\documentclass{aa}  
\usepackage{orcidlink}

\usepackage{graphicx}
\usepackage{txfonts}
\newcommand{\equ}[1]{eq.~(\ref{eq:#1})}

\newcommand{\se}[1]{\S\ref{sec:#1}}

\newcommand{\Fig}[1]{Figure~\ref{fig:#1}}

\newcommand{\tab}[1]{Table~\ref{tab:#1}}
\newcommand{\be}{\begin{equation}}
\newcommand{\ee}{\end{equation}}
\newcommand{\bea}{\begin{eqnarray}}
\newcommand{\eea}{\end{eqnarray}}

\newcommand{\msun}{{\rm M}_\odot}
\newcommand{\Msun}{M_\odot}

\newcommand{\ifm}[1]{\relax\ifmmode#1\else$\mathsurround=0pt #1$\fi}
\newcommand{\kms}{\ifmmode\,{\rm km}\,{\rm s}^{-1}\else km$\,$s$^{-1}$\fi}

\newcommand{\Mpc}{\,{\rm Mpc}}
\newcommand{\kpc}{\,{\rm kpc}}
\newcommand{\pc}{\,{\rm pc}}
\newcommand{\Gyr}{\,{\rm Gyr}}

\newcommand{\Myr}{\,{\rm Myr}}
\newcommand{\GyrI}{\,{\rm Gyr}^{-1}}

\newcommand{\ltsima}{$\; \buildrel < \over \sim \;$}
\newcommand{\lsim}{\lower.5ex\hbox{\ltsima}}
\newcommand{\gtsima}{$\; \buildrel > \over \sim \;$}
\newcommand{\gsim}{\lower.5ex\hbox{\gtsima}}

\def\Ms{M_*}
\def\Re{R_{\rm e}}

\usepackage{color}

\begin{document}

   \title{Accelerated size evolution in the FirstLight simulations from z=14 to z=5}

   \author{D. Ceverino
          \inst{1,2}, 
          Y. Nakazato \inst{3}, N. Yoshida  \inst{4,5,6}, R. S. Klessen \inst{7,8}, 
          S. C. O. Glover \inst{7}, \and L. Costantin \inst{9}
          }

   \institute{Departamento de Fisica Teorica, Modulo 8, Facultad de Ciencias, Universidad Autonoma de Madrid, 28049 Madrid, Spain\\
              \email{daniel.ceverino@uam.es} 
         \and
             CIAFF, Facultad de Ciencias, Universidad Autonoma de Madrid, 28049 Madrid, Spain
         \and
         Center for Computational Astrophysics, Flatiron Institute, 162 5th Avenue, New York, NY 10010
         \and
         Department of Physics, The University of Tokyo, 7-3-1 Hongo, Bunkyo, Tokyo 113-0033, Japan 
         \and 
         Kavli Institute for the Physics and Mathematics of the Universe (WPI), UT Institute for Advanced Study, The University of Tokyo, Kashiwa, Chiba 277-8583, Japan
         \and
         Research Center for the Early Universe, School of Science, The University of Tokyo, 7-3-1 Hongo, Bunkyo, Tokyo 113-0033, Japan
         \and
         Universit\"{a}t Heidelberg, Zentrum f\"{u}r Astronomie, Institut f\"{u}r Theoretische Astrophysik, Albert-Ueberle-Str. 2, 69120
Heidelberg, Germany
	\and
	Universit\"{a}t Heidelberg, Interdisziplin\"{a}res Zentrum f\"{u}r Wissenschaftliches Rechnen, INF 205, 69120, Heidelberg, Germany
	\and
	Centro de Astrobiología (CAB), CSIC-INTA, Ctra. de Ajalvir km 4, Torrejón de Ardoz, E-28850 Madrid, Spain \\
             }

   \date{Received Month day, year; accepted Month day, year}

\titlerunning{Size evolution in FirstLight}
\authorrunning{Ceverino et al.}

 
  \abstract
   {Galaxies grow very rapidly during the first Gyr of the Universe, mostly driven by high galaxy efficiencies, particularly relevant at $z>5$.
   This  efficiency is related to high gas densities and/or compact gas distributions within these early galaxies.}
   {We want to understand the evolution of the size of galaxies at cosmic dawn, from $z=14$ to $z=5$ and its main drivers. }
   {We use the FirstLight database of 430 zoom-in cosmological simulations and radiative transfer calculations to generate synthetic images in seven JWST bands.
  We add observational effects, inspired by recent JWST deep extragalactic surveys.}
   { The size-mass relation is already in place at $z\simeq14$ and it shows a large diversity of galaxy sizes at a fixed mass. 
    Extended (compact) galaxies tend to have higher (lower)  specific star-formation rate (sSFR). 
   The mass-dependent slope does not evolve significantly.
   This is driven by a complex interaction between stellar light and dust. Differential dust attenuation dims galaxy centers and it makes larger sizes, modifying the mass-size slope even in the rest-frame optical.
 At a fixed mass, galaxy size evolves very fast, as the normalization of the size-mass relation increases by 0.5 dex between $z\simeq14$ and $z\simeq6$, in 600 Myr.
The SFR surface density  increases with redshift, driven by
 higher  sSFRs and smaller sizes at higher redshifts.
 }   
   {Size evolution at a fixed stellar mass accelerates at cosmic dawn, driven by an increasing galaxy efficiency at  $z\geq5$.}


   \maketitle
   
      \nolinenumbers  

%

\section{Introduction}
\label{sec:intro}

Galaxies grow very rapidly during the first Gyr in the history of the Universe \citep{Stark16, Stark26}.
The James Webb Space Telescope (JWST) has observed a higher-than-expected number of bright (sometimes massive) galaxies at extremely high redshifts, $z=9-15$ \citep{Naidu22, Castellano22, Adams23, Atek23, Donnan23, Harikane23, PerezGonzalez23, Leung23, Finkelstein23, Yan23, Hainline24,  Casey24,  PerezGonzalez25}. 
This accelerated mass growth can be driven by a high galaxy efficiency at these high redshifts \citep{Dekel23}.
 \cite{PaperV} define this efficiency as the 
 ratio between galaxy growth and halo growth. 
 It increases with redshift even at a fixed halo mass due to 
 higher gas densities and the smaller sizes of star forming (SF) regions. 
Therefore, the evolution of the galaxy size could also be faster than at later times
and it may be related to this accelerated growth.

\begin{table*} 
\caption{The FirstLight suite of simulations.}
 \begin{center} 
 \begin{tabular}{cccccc} \hline 
 Box size	&  Effective resolution  & 	DM mass resolution	& minimum cell size   &  Number of galaxies & Reference \\
 \hline 
 10 Mpc/h 	   &	$2048^3$ &	$10^4 \msun$		&    8.7-17 pc	  &      201 &  \cite{PaperII}  \\ 
 20 Mpc/h		& $4096^3$ &      $10^4 \msun$		&    8.7-17 pc	  &      114  & \cite{PaperII}	\\
 40 Mpc/h       &	  $4096^3$ &      $ 8 \times 10^4 \msun$&   17-35 pc	  &       33  & \cite{PaperV}  \\	
 80 Mpc/h	     &	   $4096^3$ &     $ 6 \times 10^5 \msun$&   35-70 pc	  &       53  & This work \\
\hline 
 \end{tabular} 
 \end{center} 
 
 
 \label{tab:firstlight} 
 \end{table*} 	

Observations have shown a decrease of the galaxy effective radius, or half-light radius, with increasing redshifts.
Before JWST, observations with the Hubble Space Telescope (HST) have shown that 
the relation between size and stellar mass for SF galaxies is well described by a single power law \citep[e.g.][]{vanderWel14, Mowla19, Nedkova21}:
\begin{equation} 
{\rm log}(\Re/ \kpc) = \alpha  \ {\rm log}( \Ms /  10^9 \ \msun) + \beta \ , 
\label{eq:Refit}
\end{equation}
where $\alpha$ describes the slope of the relation and $\beta$ is its redshift-dependent normalization:
\begin{equation} 
\beta= \alpha_z  \ {\rm log} (1+z)_{z5} + \beta_{z5} \ ,
\label{eq:Revo}
\end{equation}
where $ (1+z)_{z5}=(1+z)/6$.  $ \alpha_z$ describes the redshift evolution and  $\beta_{z5}$ is the logarithm of the size of a galaxy with $\Ms = 10^9 \ \msun$ at $z=5$.
HST observations at $z\leq3$ show an important decrease of the galaxy size with redshift, $\alpha_z \simeq -1$, even when measurements differ between rest-frame UV and rest-frame optical \citep{Shibuya15}. On the other hand, the $\alpha$ slope remains positive, $\alpha \simeq0.2$, and  roughly constant in time. 
These trends continue in Lyman Break Galaxies (LBG) at higher redshifts, $z\geq3$ \citep{Mosleh12, Shibuya15}.

The advent of JWST provides hints of the size growth to extremely high redshifts \citep{Morishita24, Allen25, Danhaive25}.
Some observations show a slower evolution, $\alpha_{z} > -1$ \citep{ Morishita24, Allen25} up to $z\simeq10-14$, although the number of examples at these high redshifts is still small.  Some of these galaxies are quite compact. For example, JADES-GS-z14-0, one of the most distant galaxies observed by JWST, has a size of only 260 pc at $z=14$ \citep{Robertson24}.
A recent sample of objects at  $z=10-16$ drives a steeper evolution, $\alpha_{z}  \simeq -1.5 $ \citep{Ono25}.
On the other hand, the number and quality of the  observations at slightly lower redshifts $z=5-7$ has significantly increased. This yields a size-mass relation with a notable intrinsic scatter.   
The origin of this scatter as well as the size evolution at $z>10$ remain poorly understood.

Galaxy size at high-z is heavily influenced by wet compaction \citep{DekelBurkert}. This process is driven by a rapid loss of angular momentum (and energy) that generates a fast inflow of gas to the  galaxy center. If the time-scale of this inflow is faster than the star-formation time-scale, most of this gas fuels a central starburst that reduces the effective radius of the galaxy \citep{Zolotov15}.
This process is especially relevant in galaxies above a stellar mass of $10^{8.5} \ \msun$ at cosmic dawn \citep{Cataldi25}.
At lower masses, stellar feedback counterbalances these inflows.
\cite{Lapiner23} found that 
mergers are the main source of compaction. 
Mergers are  common at cosmic dawn and we expect major mergers to happen every 100 Myr on average, especially in massive galaxies, consistent with recent observations \citep{Calabro26}. Other   
secondary sources includes counter-rotating stream collisions and violent disk instabilities. 

Modern cosmological simulations of first galaxies encounter difficulties to reproduce the size-mass relation and/or its evolution \citep{Costantin23, Roper22, LaChance25, Shen24, McClymont25}.
Most of them predict flat or negative slopes, $\alpha\leq0$, due to a complex interplay between the effects of spatial resolution and dust attenuation.
Compact galaxies with small, sub-kpc sizes are difficult to reproduce.
Measurements of the half-light radius are very sensitive to how dust and stars are mixed. For example, high dust column densities at the galaxy center may decrease the light coming from a proto-bulge and yield a larger size compared to the intrinsic stellar distribution \citep{Costantin23}.
Different stellar populations may also have different spatial distributions. For example, a galaxy may host a young central starburst  embedded in an extended and mature population.  
Therefore, the rest-frame UV size, tracing the young population, can be smaller than the size in the rest-frame optical.
 Dust attenuation is also sensitive to wavelength and this complicates the interpretation of the observed sizes.
Therefore, models should be able to generate synthetic images that could be directly compared to observations at different bands.


The FirstLight (FL) simulations \citep{PaperI} provide the largest suite of zoom-in simulations ($\sim400$ independent regions) with the highest resolution (up to $\sim10$ proper pc) to date.
These numerical experiments are suitable for a statistical study of the size growth from extremely high redshifts, $z\simeq14$, to the end of Reionization, $z\simeq5$. Our goal is also to unveil the origin of the scatter in the size-mass relation and 
to connect the size evolution with the evolution of the galaxy efficiency and the SFR surface density.
The outline of this paper is as follows.
Section \se{runs} summarizes the FirstLight simulations. 
Section \se{skirt} describes the methodology used to generate synthetic images.
Section \se{results} provides the main findings, Section \se{discussion} discusses the results and Section \se{conclusions} provides a summary and a general conclusion.

\section{The FirstLight simulations}
\label{sec:runs}
    
The FirstLight simulations are multi-object, zoom-in cosmological simulations of the first galaxies up to $z\simeq5$.
The initial conditions of a mass-complete sample of galaxies are first described by  \cite{PaperI}.
The current suite is composed of four cosmological boxes, summarised in  \tab{firstlight}.
The 10-Mpc/h and 20-Mpc/h samples contain all halos within these volumes with a maximum circular velocity, V$_{\rm max}>50 \kms$ and  V$_{\rm max}>100 \kms$, respectively, at $z=5$.
Similarly, the 40-Mpc/h box contains galaxies with  V$_{\rm max}>180 \kms$ and the 80-Mpc/h box has V$_{\rm max}>250 \kms$, containing the most massive galaxies of the sample.
Overall, the  FirstLight database includes 430 galaxies, named FLxxx, where xxx denotes the label of the galaxy in the initial conditions.
Although the last two boxes have lower resolution, no resolution effects have been reported in previous papers \citep{PaperV}.
 
The simulations are performed with the  $N$-body+Hydro \textsc{ART} code
\citep{Kravtsov97,Kravtsov03, Ceverino09, Ceverino14, PaperI}.
Gravity and hydrodynamics are solved by an Eulerian, adaptive mesh refinement (AMR) approach.
The code includes  astrophysical processes relevant for galaxy formation, such as gas cooling by hydrogen, helium and metals.
Photoionization heating uses a redshift-dependent cosmological UV background with partial self-shielding. 

Star formation and feedback (thermal+kinetic+radiative) models are described in \cite{PaperI}.
In short, star formation is assumed to occur at densities above a threshold of 1 cm$^{-3}$ and at temperatures below $10^4$ K. The code implements a stochastic star formation model that scales with the gas free-fall time \citep{Schmidt, Kennicutt98}.
In addition to thermal energy feedback, the simulations use radiative feedback, as a local approximation of radiation pressure. This model adds non-thermal pressure to the total gas pressure in regions where ionizing photons from massive stars are produced and trapped. The model of radiative feedback is named RadPre\_IR in \cite{Ceverino14} and it uses a moderate trapping of infrared photons. 
The kinetic feedback model also includes the injection of momentum coming from the (unresolved) expansion of gaseous shells from supernovae and stellar winds \citep{OstrikerShetty11}.  More details can be found in \cite{PaperI, Ceverino14, Ceverino10} and in  \cite{Ceverino09}. 
The simulations follow metals from  supernovae type-II and type Ia, using yields that approximate the  results from \cite{WoosleyWeaver95}, as described in  \cite{Kravtsov03}.  
 
 This database has been used in several studies: merger-driven clumps \citep{Nakazato24}, mini-quenching  \citep{Dome24}, and galaxy efficiency \citep{PaperV}.
Other papers using this database predict rest-frame spectral energy distributions (SEDs), and emission from optical  \citep{PaperIII, PaperIV} and far-infrared \citep{Nakazato23} lines.  
 In addition, FirstLight predicts
 bursty SF galaxies \citep{PaperII} and
 a weak evolution of the mass-metallicity relation at $z\ge5$ \citep{Langan20}, consistent with current findings \citep{Curti23, Venturi24}.
 Dust masses are consistent with other models \citep{Mushtaq23} and \cite{Nakazato26} describe spatially resolved dust properties.
 Finally, \cite{Cataldi25} focus on compaction at high-$z$.

\section{Generation of synthetic images with SKIRT}
\label{sec:skirt}

   \begin{figure*}
   \centering
   \includegraphics[width=2 \columnwidth]{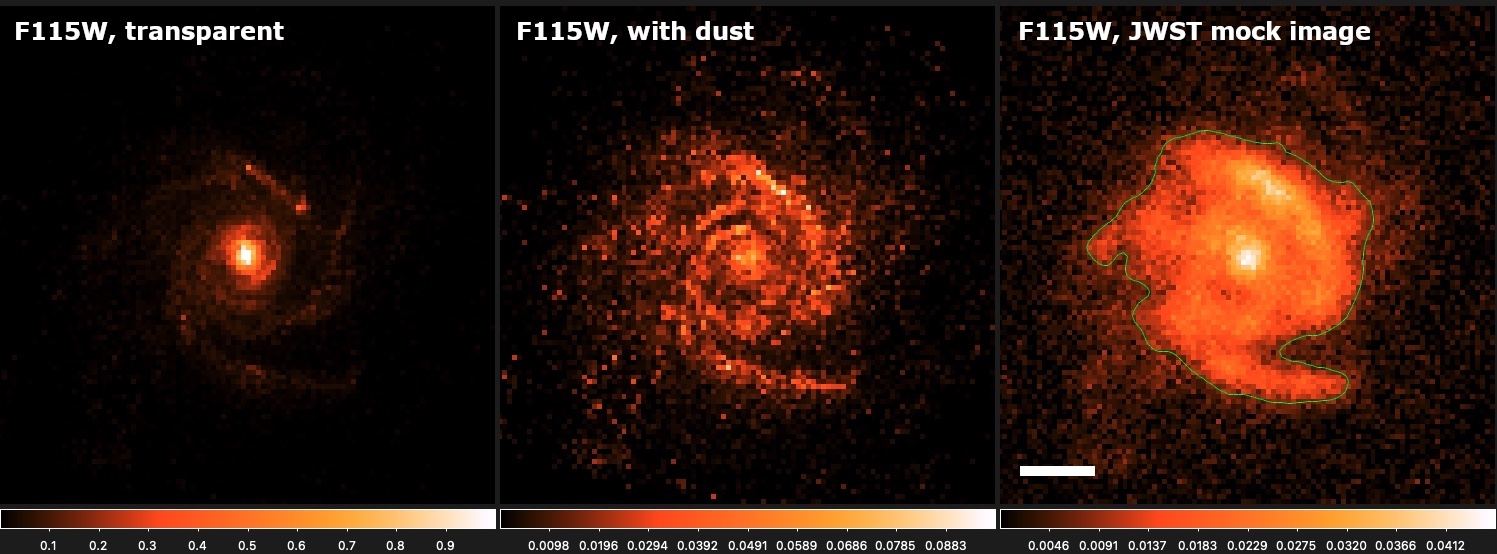}
   \includegraphics[width=2 \columnwidth]{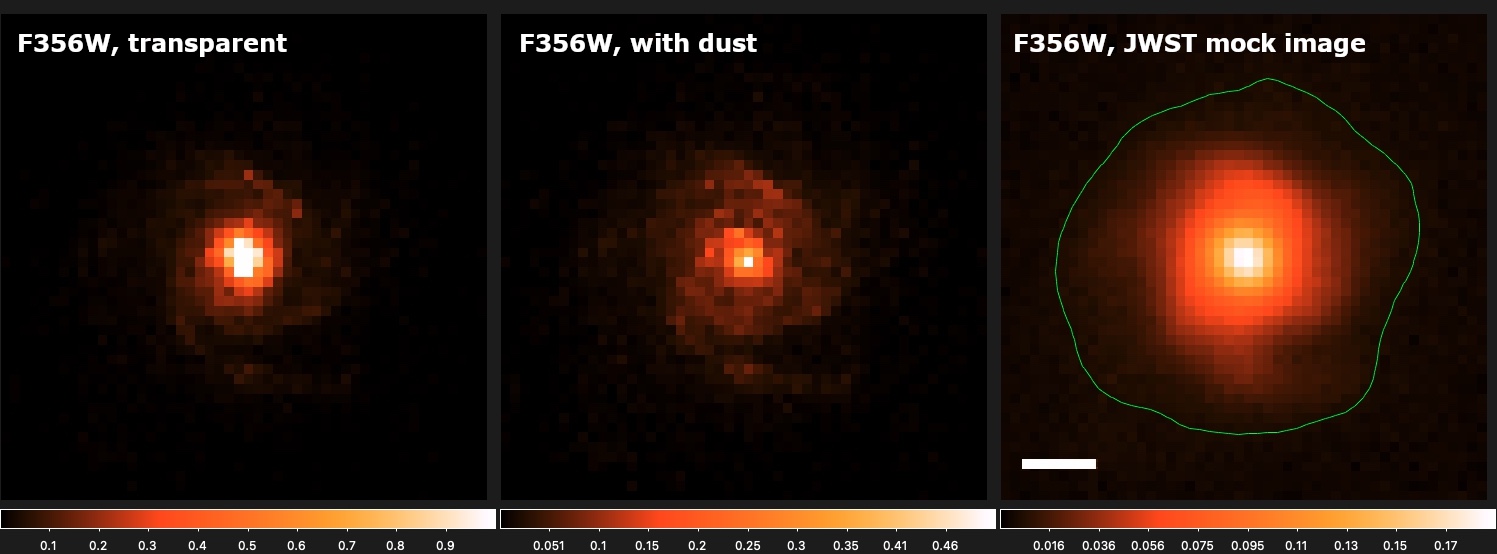}

   \caption{Example of an extended, high-mass disk at $z=6$ (FL961). Top panels show a short-wavelength band (F115W) and bottom panels show a longer wavelength (F356W). Left panels represent the transparent case. Middle panels include  absorption and scattering by dust. Right panels add instrumental effects. 
 An effective radius of $\Re \simeq1.5 \kpc$  (horizontal bar) is computed using the area that contains half of the luminosity of a region inside a given isophote (green contour).
   Each image shows a face-on view with a side length of 10 kpc. Color bars show surface brightness in ${\rm MJy \ sr}^{-1}$.}  
               \label{fig:example1}%
    \end{figure*}          
   \begin{figure*}
   \centering
   \includegraphics[width=2 \columnwidth]{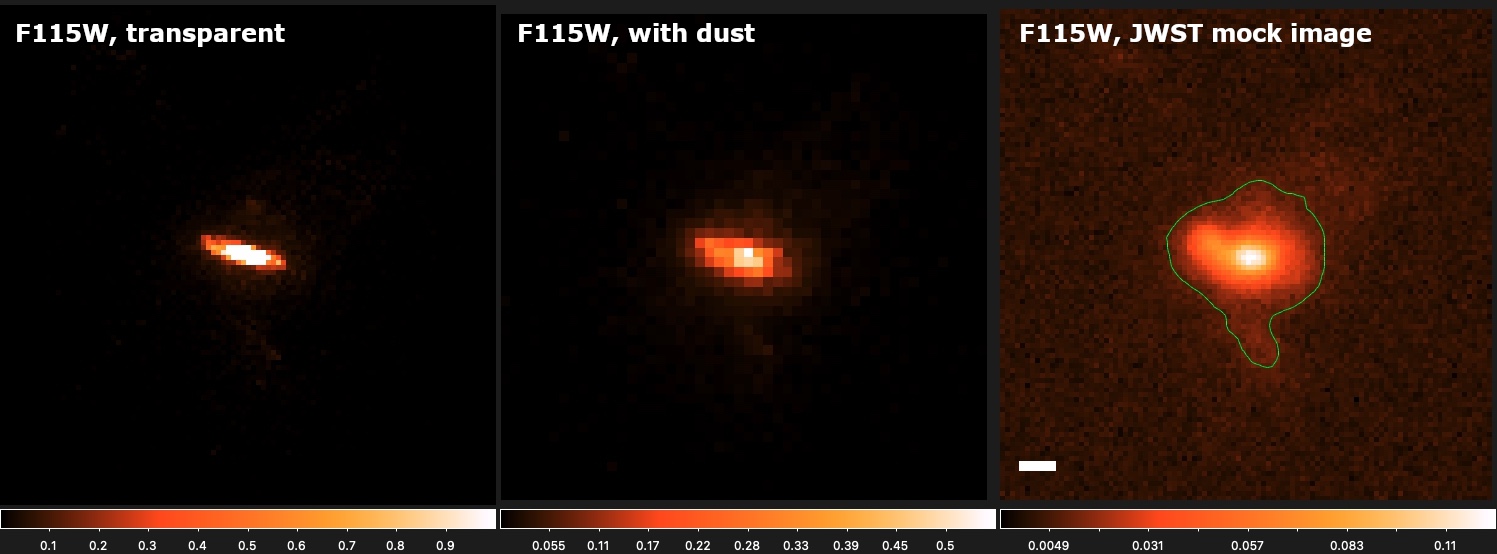}

   \caption{Example of a compact, high-mass disk at $z=6$ (FL964) with an effective radius of $\Re \simeq0.7 \kpc$ in the rest-frame UV (F115W). Each image shows an edge-on view with a side length of 10 kpc. Labels are defined as in \protect\Fig{example1}.} 
               \label{fig:example2}%
    \end{figure*}  
   \begin{figure*}
   \centering
   \includegraphics[width=2 \columnwidth]{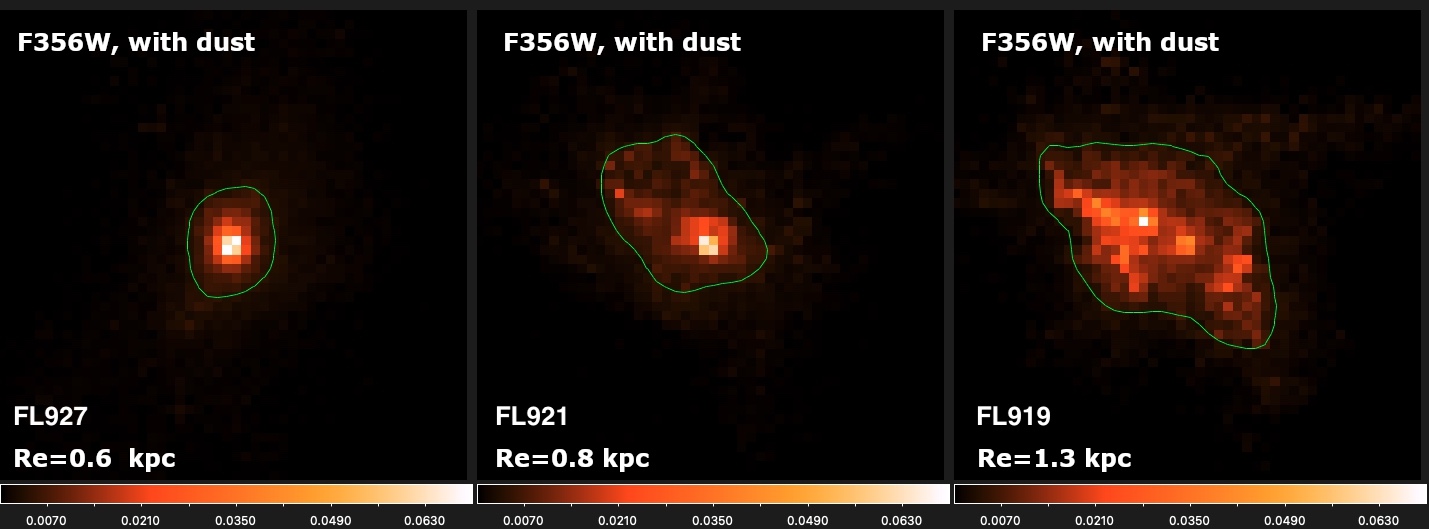}
  
   \caption{Three examples of low-mass galaxies with $\Ms\simeq10^9 \ \msun$ at $z=6$ in the F356W band with a compact (left), average (center), and extended (right) morphology. Each image has a side length of 10 kpc. Labels are defined as in \protect\Fig{example1}. }
               \label{fig:example4}%
    \end{figure*}  

   \begin{figure*}
   \centering
   \includegraphics[width=2 \columnwidth]{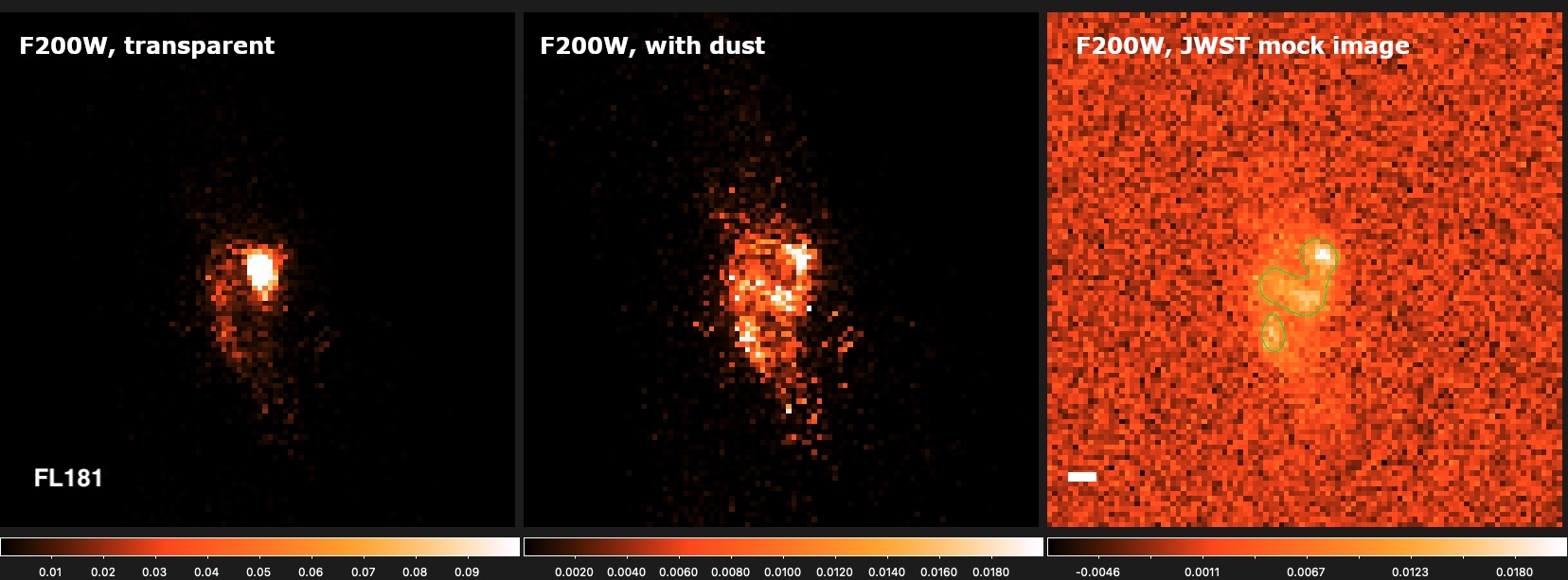} 
   \includegraphics[width=2 \columnwidth]{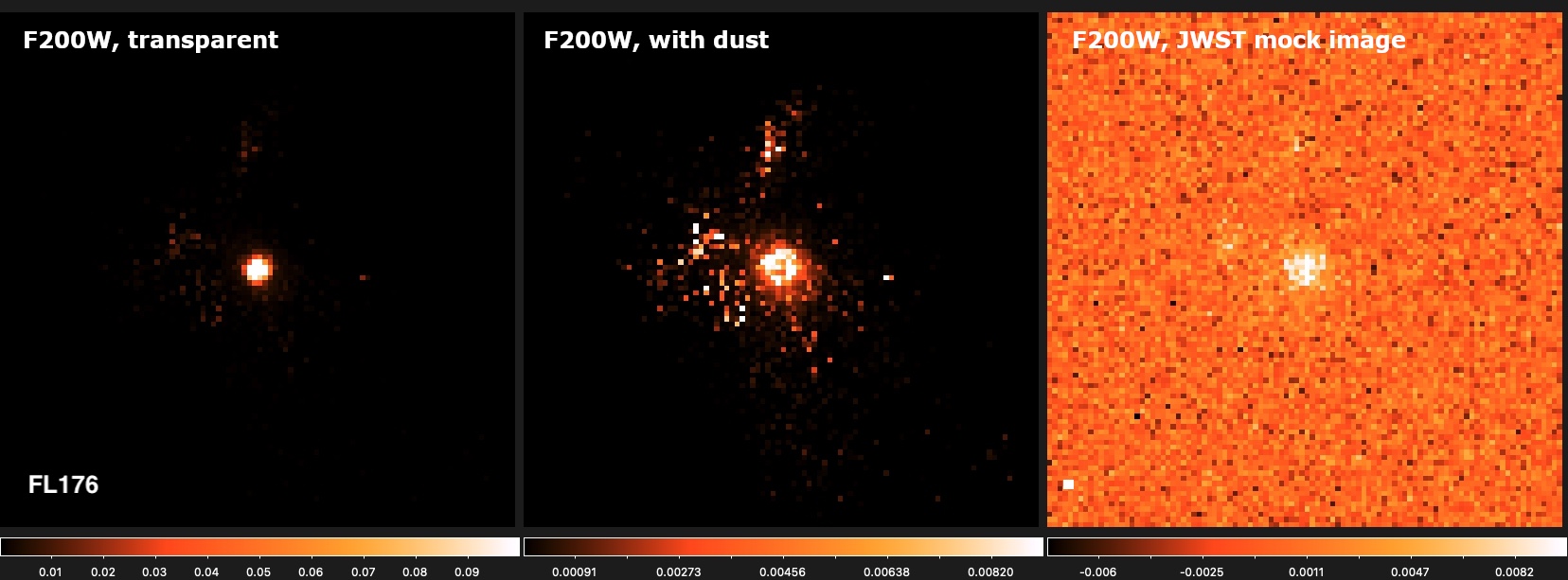}

   \caption{Two examples of low-mass galaxies with $\Ms\simeq10^9 \ \msun$ at $z=10$ in the F200W band. Top panels show a relatively extended galaxy with $\Re \simeq 0.5 \kpc$, undergoing a merger. Bottom panels show a compact counterpart with  $\Re \simeq 0.16 \kpc$, similar to observed galaxies like JADES-GS-z14-0 \protect\citep{Carniani24}. Each image has a side length of 10 kpc. Labels are defined as in \protect\Fig{example1}. }
               \label{fig:example3}%
    \end{figure*} 

The comparison between size measurements in observations and simulations requires synthetic images that take into account the dust-light interactions as well as the instrumental features of a real observation \citep{Snyder15, Snyder15b, Trayford17, Popping22, Ikhsanova25}.

We use the publicly available code SKIRT \footnote{Version 9, http://www.skirt.ugent.be}  \citep{CampsBaes15, Verstocken17, CampsBaes20}
to produce noiseless synthetic images of the FirstLight galaxies. 
For each snapshot, we first compute the intrinsic stellar emission using the BPASS population synthesis code \citep{Eldridge17}, assuming a \cite{Chabrier03} IMF with an upper mass of 300 $\Msun$. Nebular emission is computed using TODDLERS \citep{Kapoor23}, assuming a star formation efficiency, SFE=0.1, and a constant HII density of 100 cm$^{-3}$.
The conditions of HII regions at these high redshifts are poorly constrained and these are the same parameters used in a previous paper \citep{PaperIII}.
A more detailed modelling of nebular emission can be found in \cite{Nakazato23, Nakazato24}.
The effect of the emission lines can be quantified by comparing the size of FL galaxies at $z=5.25$, measured at two different bands (F356W and F277W).
At that redshift, only the latter band includes strong emission lines.
The size difference is about 10\%.
The simulated equivalent widths are typically less than $1000 \AA$ \citep{PaperIII}.
This feature plus the relatively wide bands used in this work yield a small effect of  the nebular emission on the size of galaxies.

The distribution of dust is computed from the density of metals, assuming a constant dust-to-metal ratio of DMR=0.4 \citep{Snyder15, Behrens18, Lovell21}. 
This assumption may not work at low metallicities \citep{Popping17}, although this regime corresponds to low-mass galaxies that are too faint to be included in our study. 
For galaxies more massive than $\Ms \geq10^7 \ \msun$, our estimates of dust mass are consistent with previous models within the observed scatter \citep{Mushtaq23}.
The size distribution and composition of the dust determine its interaction with photons.
However, the properties of dust at these high redshifts are poorly constrained.
We use a \cite{WeingartnerDraine01} dust mix, appropriate for a Milky-Way extinction curve.
Assuming a SMC dust mix generates very similar size measurements, a less than 8\% difference on average.

Using SKIRT, we compute the radiative transfer of the intrinsic photons throughout the distribution of dust and generates a synthetic image. We select a fixed field-of-view of 10 kpc and $100 \times 100$ pixels for the short wavelength channel and $50 \times 50$ pixels for the long wavelength channel.
The  quality of these calculations depends on  the number of photon packets. After several tests, we use 10$^6$ packets per image. This number is sufficient for the selected number of pixels. Higher numbers (up to 10$^9$ packages) do not significantly improve these images. 
At each available snapshot, we use 7 JWST wide-bands (from F090W to F444W) and 9 different orientations, including face-on and edge-on views, as well as randomly selected angles. The complete database of 54200 images at each redshift will be publicly available \footnote{http://odin.ft.uam.es/FirstLight/index.html}.
Different orientations do not  change the size determinations by more than 1\% on average.
This implies that the majority of galaxies at these high redshifts are triaxial or spheroidal systems \citep{Ceverino15b}.

The estimation of the galaxy size and its comparison with observations requires the inclusion of a background noise and the effects of the telescope point-spread-function (PSF) \citep{Ma18}. 
We approximately mimic the background noise levels of the JADES-GDS survey  \citep{Eisenstein23, Tacchella23, Robertson24}, with an aproximated 5$\sigma$-limiting magnitude (in ABmag) of 30. We model the background by using a gaussian noise centred at zero. We select its dispersion based on the standard deviation of the noise in the source-less fields of the same survey. We use $\sigma=0.002 \ {\rm MJy \ sr}^{-1}$ for F115W, F150W, F200W bands and  $\sigma=0.001 \ {\rm MJy \ sr}^{-1}$ for F277W, F356W, F444W. The PSF is also modeled as a gaussian filter with a standard deviation of 0.031 arcsec for the short-wavelength channels and 0.063 arcsec for the long-wavelength channels. 
A more sophisticated model of the instrumental effects, including the available JWST PSF, is not necessary for the estimation of the size of galaxies \citep{Ma18}.
Studying tidal features and other morphological substructures requires a more detailed modelling \citep{Mantha19}.
This is beyond the aims of the present paper.

The computation of the half-light radius uses a non-parametric, pixel-based method \citep{Ribeiro16, Ma18, Marshall22}. 
This method is preferred over other standard approaches, like Sersic fitting, because high-$z$ galaxies are quite irregular and clumpy \citep{Ma18}. 
Parametric methods assume smooth and symmetric models that are not good fits of these images. 
On the other hand, our pixel-based method assumes a zero flux below a surface brightness limit that is four times higher than the noise ($4 \sigma$), following \cite{Ma18}. We compute the effective radius using the area that contains half of the luminosity above this surface brightness limit, $ R_{\rm e}= ( A_{50}/ \pi )^{0.5}$.
Other variables, like the galaxy stellar mass ($\Ms$), and the star-formation-rate (SFR), are taken from the FirstLight database \citep{PaperII, PaperIV} and therefore these results can be compared with previous papers.

\section{Results}
\label{sec:results}

\subsection{Examples of compact and extended galaxies} 
\label{sec:examples}
    
The synthetic images from the FirstLight simulations show a large diversity of galaxies.
As a first example, \Fig{example1} shows a face-on view of a high-mass galaxy ($ \Ms \simeq 2 \times 10^{10} \ \msun$)  at $z=6$ in two bands (F115W and F356W). The first  band covers the rest-frame UV and the second band corresponds to the rest-frame optical.
The image in the rest-frame UV without dust attenuation shows a  grand-design spiral pattern with a strong starburst at the centre (${\rm SFR}=60 \ \msun {\rm yr}^{-1}$).
The inclusion of dust severely attenuates the UV light from the central starburst in comparison with the spiral pattern. As a result, the effective radius increases due to the stronger attenuation of the disc center. 
The addition of instrumental effects broadens these features but it does not affects the size determination. 
The galaxy disc is extended, with a radius of  $\Re\simeq 1.5 \kpc$.

Dust also modifies the size measurements in the rest-frame optical (bottom panels of  \Fig{example1}). 
Without dust, the luminosity is again dominated by the central burst. Therefore, the effective radius shrinks to 0.7 kpc. 
The effect of dust diminishes the luminosity at the center and the light from the stellar disc is more relevant
The effective radius increases to 1.5 kpc, very similar to the value in the rest-frame UV.
The PSF at this band is also broader and many disc features disappear, but the observational effects do not change the size measurement.

FirstLight galaxies have a large diversity of sizes, even between galaxies at similar mass and redshift. \Fig{example2} shows a massive but compact disk with  $\Re\simeq 0.7 \kpc$ and a similar mass and SFR than the previous example. In this case, most of the intrinsic light is concentrated in the small, edge-on disk, heavily attenuated by dust. Therefore, dust attenuation is a crucial ingredient in the comparison between observations and synthetic images, especially at the massive end.        

At lower masses,  $\Ms\simeq10^9 \ \msun$ at $z=6$, the diversity of sizes is higher.  \Fig{example4} depicts three examples  in the rest-frame optical (F356W).     
 The middle panel shows a typical example ($\Re \simeq 0.8 \kpc$) of a galaxy on the star-formation main-sequence (FL921) with ${\rm sSFR}=6.4 \Gyr^{-1}$ at $z=6$ \citep{PaperII}. The galaxy exhibits an off-centered clump plus an extended region. 
 The size diversity in FirstLight galaxies correlates with galaxy properties, particularly the sSFR.
 Galaxies well above the star-formation main-sequence (0.5 dex and above) experience SF bursts, typically driven by mergers \citep{PaperII}. 
 The right image in \Fig{example4} is taken  after the first passage of a 1:6 merger (FL919). After this time, there is an extended SF region with sSFR$=17 \Gyr^{-1}$ that leads to a large galaxy with  $\Re \simeq 1.3 \kpc$ even at the rest-frame optical. 
 This is consistent with the results from simulations of galaxy mergers, where galaxies are not completely mixed and yield an extended SF region before the final coalescence \citep{Perret14}. 

On the other hand, galaxies below the star-formation main-sequence  (-0.5 dex and below) tend to be more compact.
The example in the left-panel of  \Fig{example4} (FL927) has ${\rm sSFR}=1.4 \Gyr^{-1}$ and  $\Re \simeq 0.6 \kpc$.  The image is taken $\sim100 \Myr$ after a strong burst with ${\rm sSFR}=25 \Gyr^{-1}$. \cite{PaperIV} shows its SF history.   
Well after the burst, SF gas is mostly concentrated at the centre, yielding a smaller-than-average galaxy.
If there is no substantial  gas accretion at later times, this example could be the progenitor of compact quiescent galaxies at $z=1.5-3$ \citep{Barro13}.

 This size diversity extends to higher redshifts, $z\geq10$, although galaxies at these redshifts are more compact than low-z counterparts with the same mass.
 \Fig{example3} shows two examples of low-mass galaxies,  $\Ms\simeq10^9 \ \msun$, at $z=10$.
 Galaxies are in general much smaller at high-$z$, from $\Re \simeq 0.5 \kpc$ (top row) to  $\Re \simeq 0.16 \kpc$ (bottom row).
 Dust attenuation is still relevant at these high-$z$, especially at the galaxy centers. 
This diversity of galaxy sizes agrees with recent observations at similar redshifts, such as Gz9p3 with $\Re \simeq 0.1 \kpc$ at $z=9$ \citep{Boyett24}, or  JADES-GS-z14-0 with  $\Re \simeq 0.25 \kpc$ at $z=14$ \citep{Carniani24}.

   \begin{figure*}
   \centering
   \includegraphics[width= \columnwidth]{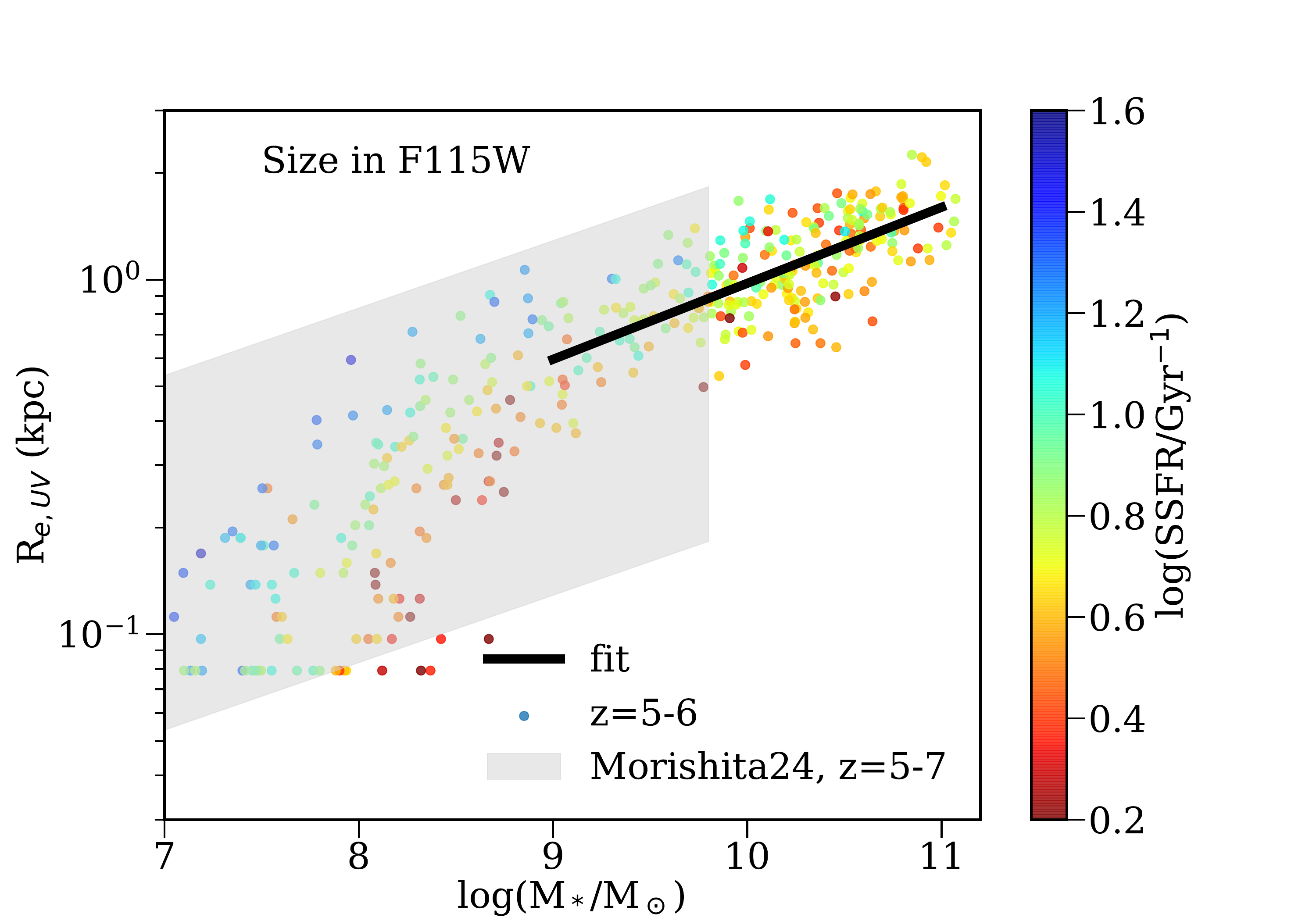}
   \includegraphics[width= \columnwidth]{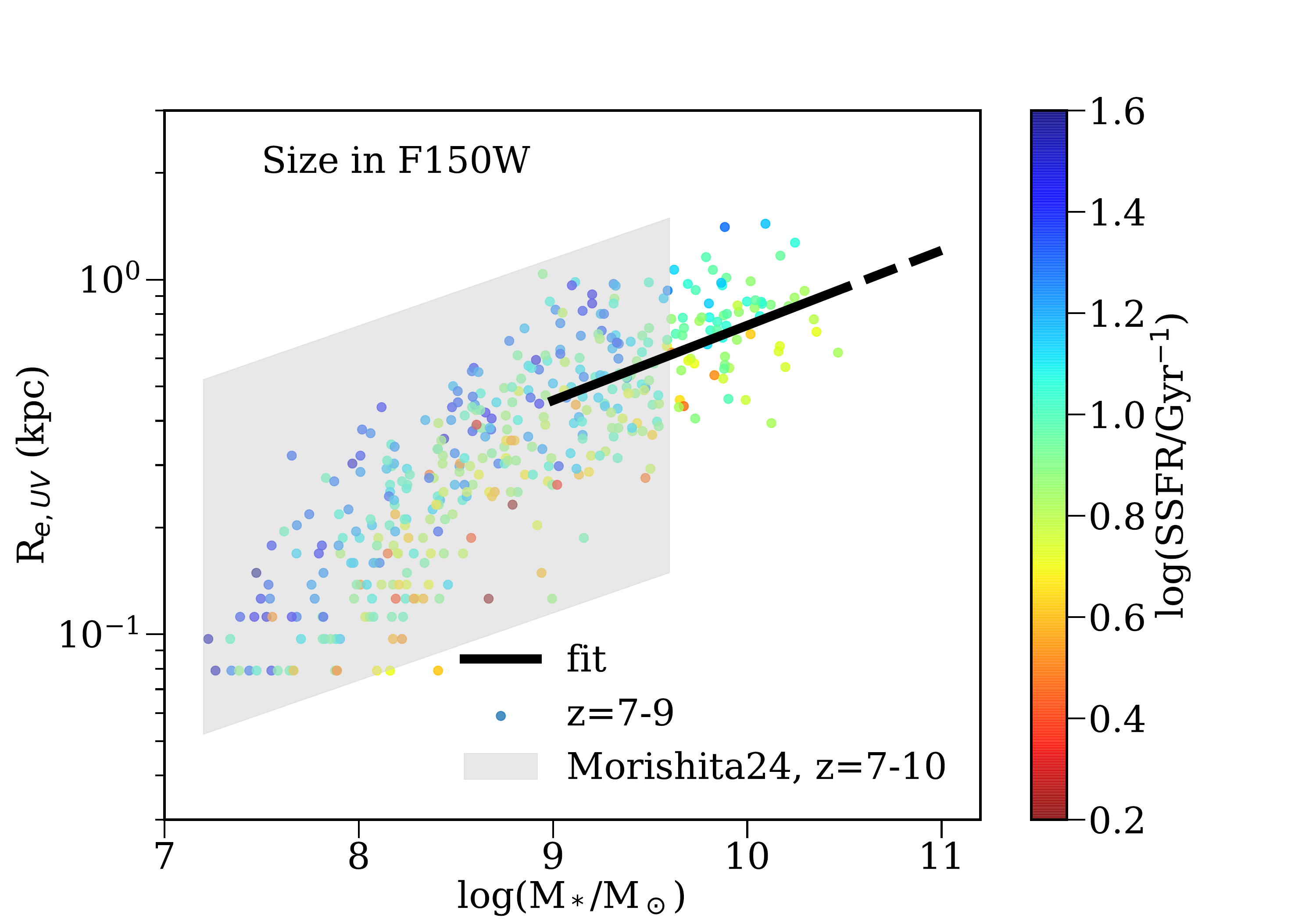}
    \includegraphics[width= \columnwidth]{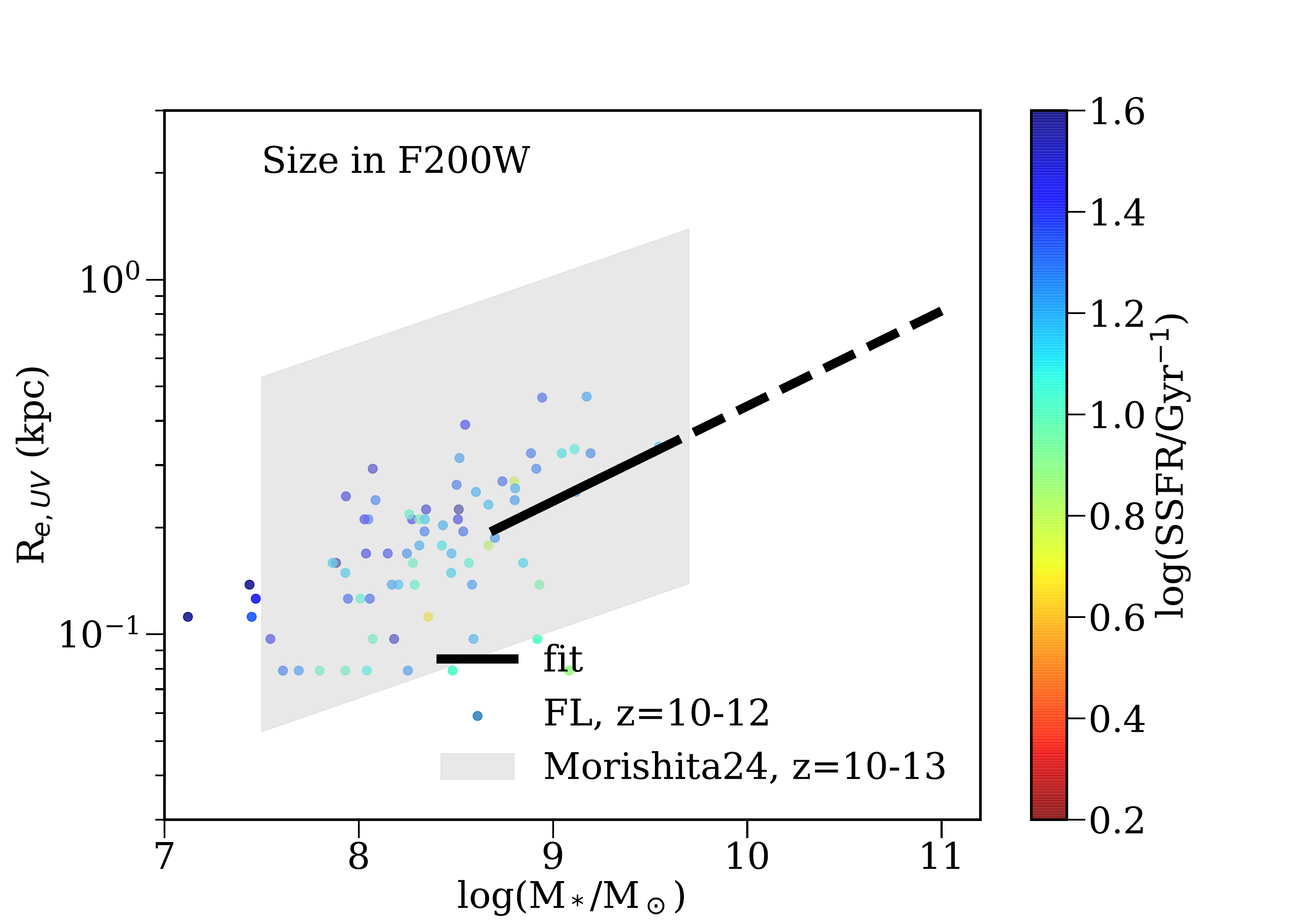}
   \includegraphics[width= \columnwidth]{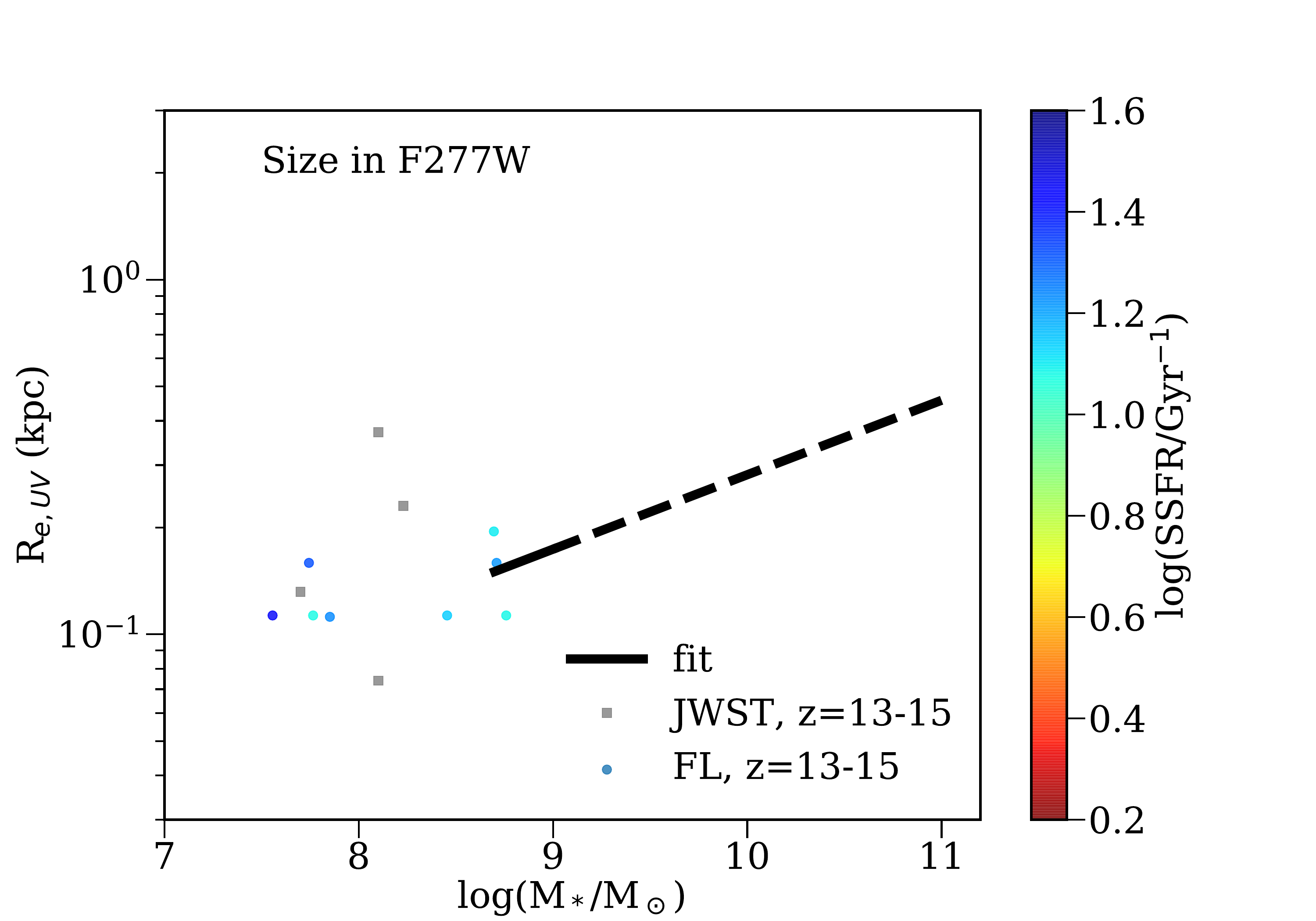}
   \caption{Size-Mass relation in the rest-frame UV at different redshifts. 
     Points are coloured by the galaxy specific star-formation-rate (sSFR). Extended (compact) galaxies tend to have higher (lower) than average sSFR.
     The lower limit in effective radius is set by the pixel size or the PSF.
   Lines represent a fit to these points above $\Ms>5 \times 10^8 \ \msun$ (\protect\tab{fit}). 
   Dash lines mark extrapolations above the sample. The slope of these lines is consistent with observations.
   Grey regions \protect\citep{Morishita24} and points \protect\citep{Morishita24,Naidu25, Donnan26} correspond to JWST results. 
  }
              \label{fig:sizeUV}%
    \end{figure*}

\subsection{Evolution of the size-mass relation in the rest-frame UV}
\label{sec:sizemassUV}

The effective radius computed in images at the rest-frame UV is sensitive to the distribution of  young and massive stars which trace the SF regions within galaxies, modulated by the effect of dust attenuation.
\Fig{sizeUV} demostrates that the size-mass relation is already in place at high redshifts, $z=5-15$. 
Galaxies with higher stellar masses have also higher rest-frame UV sizes.

The distribution of FL galaxies exhibits two different regimes above and below a stellar mass of $10^ {8.5} \ \msun$. 
Below this mass, the relation seems steeper, but  this is due to the background noise.
The size-mass relation computed using images without noise shows a similar gentle slope at high and low masses (see Appendix), although the scatter is higher at low masses due to galaxy burstiness. 

The size-mass relation at  all redshifts displays a large scatter. Some galaxies are quite compact, $\Re=100 \pc$, while others are more extended, reaching  $\Re \simeq 1 \kpc$ for the same galaxy mass.
This scatter is not random. Galaxies larger than the average at a given mass tend to have higher sSFR. These extended galaxies are above the SF main-sequence and they have higher gas fractions \citep{PaperII}. Extreme examples, significantly above the average sSFR, are usually undergoing minor mergers \citep{PaperII}. These interactions increase the rest-frame UV sizes.
On the other hand, galaxies with lower than average sSFR tend to be smaller in the rest-frame UV bands because SF gas is more concentrated at the galaxy centre. Some quiescent examples show mini-quenching events \citep{Dome24} with low gas fractions \citep{PaperIII}.

This size diversity is consistent with JWST observations \citep{Morishita24}.
However, the FL population lacks relatively massive ($\Ms \simeq 10^9 \ \msun$) and compact ($\Re\simeq200 \pc$), galaxies at $z=5-6$. They could be examples of a very short and intense, dust-free starburst in the galaxy center \citep{Ferrara25}. 
This kind of central burst makes a galaxy smaller than average in the rest-frame UV during a short period of time. Continuous gas accretion can recover the normal size growth.
These relatively rare cases are not seen in the sample of  63 random galaxies with masses between $\Ms = 10^9 - 10^{10} \ \msun$ at $z=5-6$.
At the same time, our pipeline misses low-mass ($\Ms \simeq 10^7 \ \msun$) and extended galaxies ($\Re\simeq500 \pc$). 
These are galaxies with a low surface-brightness below the minimum isophote used to compute the effective radius.
Our flux-limited sample misses about half of the galaxies with ${\rm log}(\Ms\ / \msun)=7-8$ in the F356W band at $z=5-6$ due to this selection effect. Therefore, the specific star formation rate of this low-mass population is significantly lower (sSFR=5.3 $\GyrI$) than the average value of the complete mass-selected sample (sSFR=7.3 $\GyrI$).
This effect preferentially selects compact galaxies with lower gas fractions at the low mass end.

Above a stellar mass of about $10^ 9 \ \msun$, the slope of the relation is more gentle (\tab{fit}) and the scatter is smaller.
We fit this relation using  a standard parametrization, \equ{Refit}.
We found a similar, positive slope at all redshifts ($\alpha\simeq0.21$) consistent with observed values, such as $\alpha=0.19\pm0.03$   \citep{Morishita24},  or $\alpha=0.25\pm0.06$ \citep{Danhaive25}, despite the latter being observed at slightly different wavelengths. On the other hand, \cite{Allen25} report a significantly higher value, $\alpha=0.30 \pm 0.03$, although the uncertainty is high due to the significant scatter in the observed sizes. 

The size-mass relation evolves with redshift, particularly its normalization value.
FirstLight is a volume and mass-limited sample at $z=5$. Therefore the simulations start to miss massive galaxies at $z=7-12$
because simulated volumes are too small for these rare galaxies. Despite of this, the slope remains constant within the estimated errors 
from $z=5$ to $z=12$.
The normalization significantly decreases by $\sim0.5$ dex between $z\simeq5.5$ and $z\simeq14$.
Due to the lack of massive galaxies, this last normalization is computed using the average size of galaxies with $\Ms \simeq 5 \times 10^8 \ \msun$ at $z=13-15$, keeping a fixed slope.
These compact sizes of about 100-200 pc are consistent with observed galaxies at these high redshifts, like JADESGDS-30934, spectroscopically confirmed at z = 13.2 by \cite{CurtisLake23},  JADES-GS-z14-0 \citep{Carniani24}, PAN-z14-1 \citep{Donnan26}, or MoM-z14 with $\Re \simeq 100 \pc$ at z=14.44 \citep{Naidu25}.

\subsection{Evolution of the size-mass relation in the rest-frame optical}

   \begin{figure*}
   \centering
   \includegraphics[width= \columnwidth]{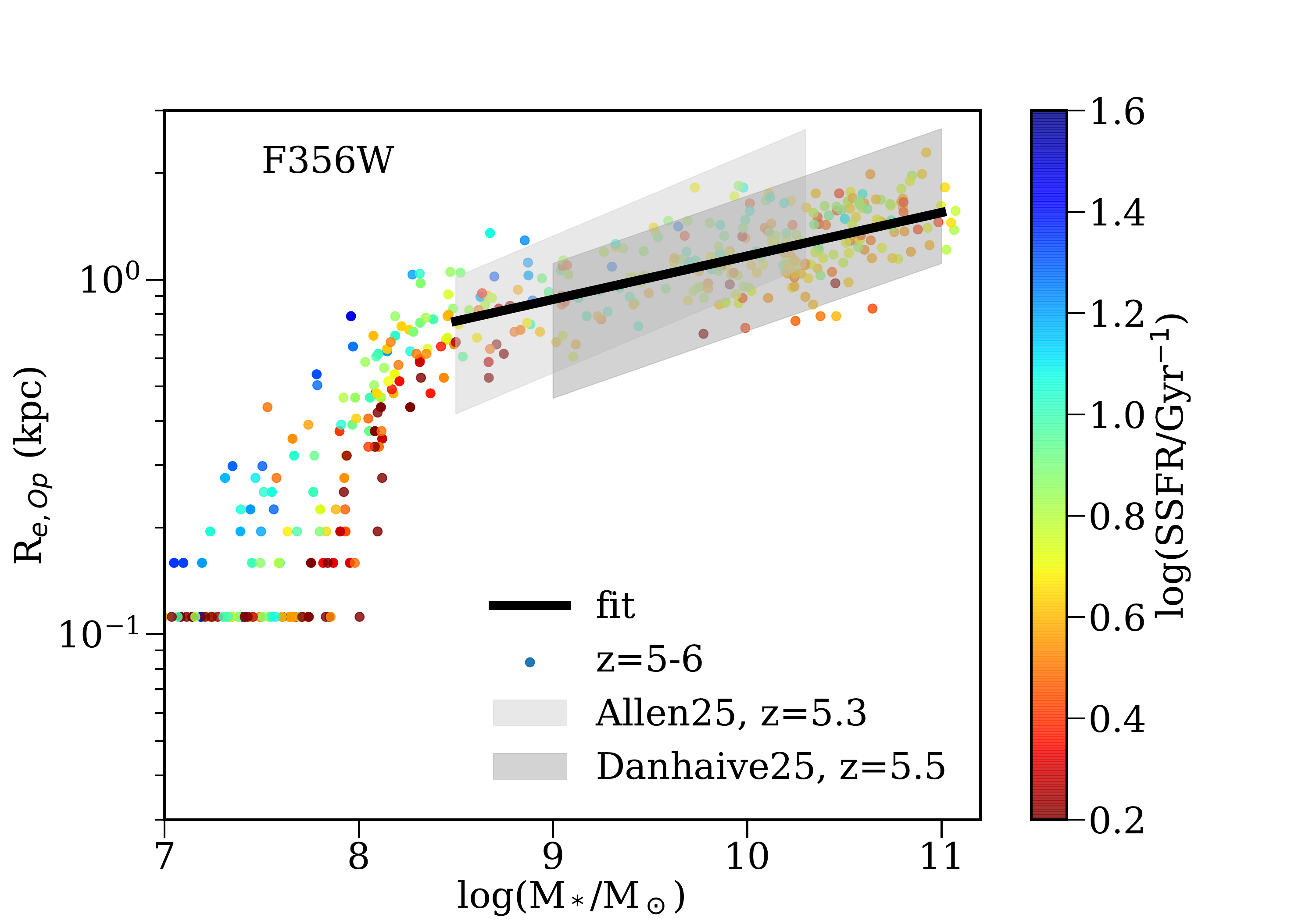}
   \includegraphics[width= \columnwidth]{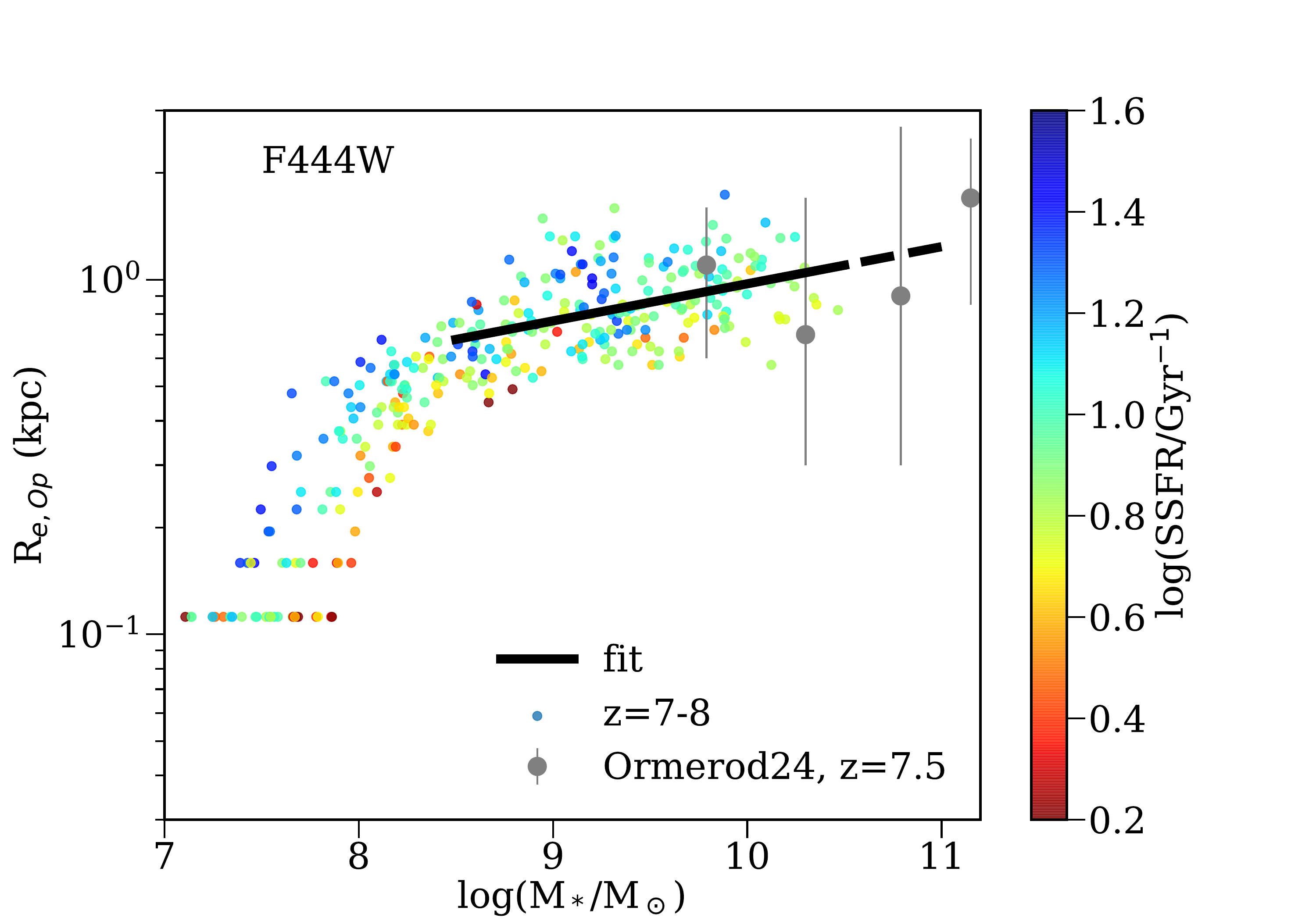}
   \caption{Size-Mass relation in the rest-frame optical at different redshifts. Points are coloured by the galaxy specific star-formation-rate (sSFR). Solid lines represent a fit to these points above $\Ms>5 \times 10^8 \ \msun$ (\protect\tab{fit}). The slope is shallower than in the rest-frame UV. Grey regions and points with error bars correspond to JWST results \protect\citep{Ormerod24,Allen25, Danhaive25}.}
              \label{fig:sizeOp}%
    \end{figure*}  
   \begin{figure*}
   \centering
   \includegraphics[width= \columnwidth]{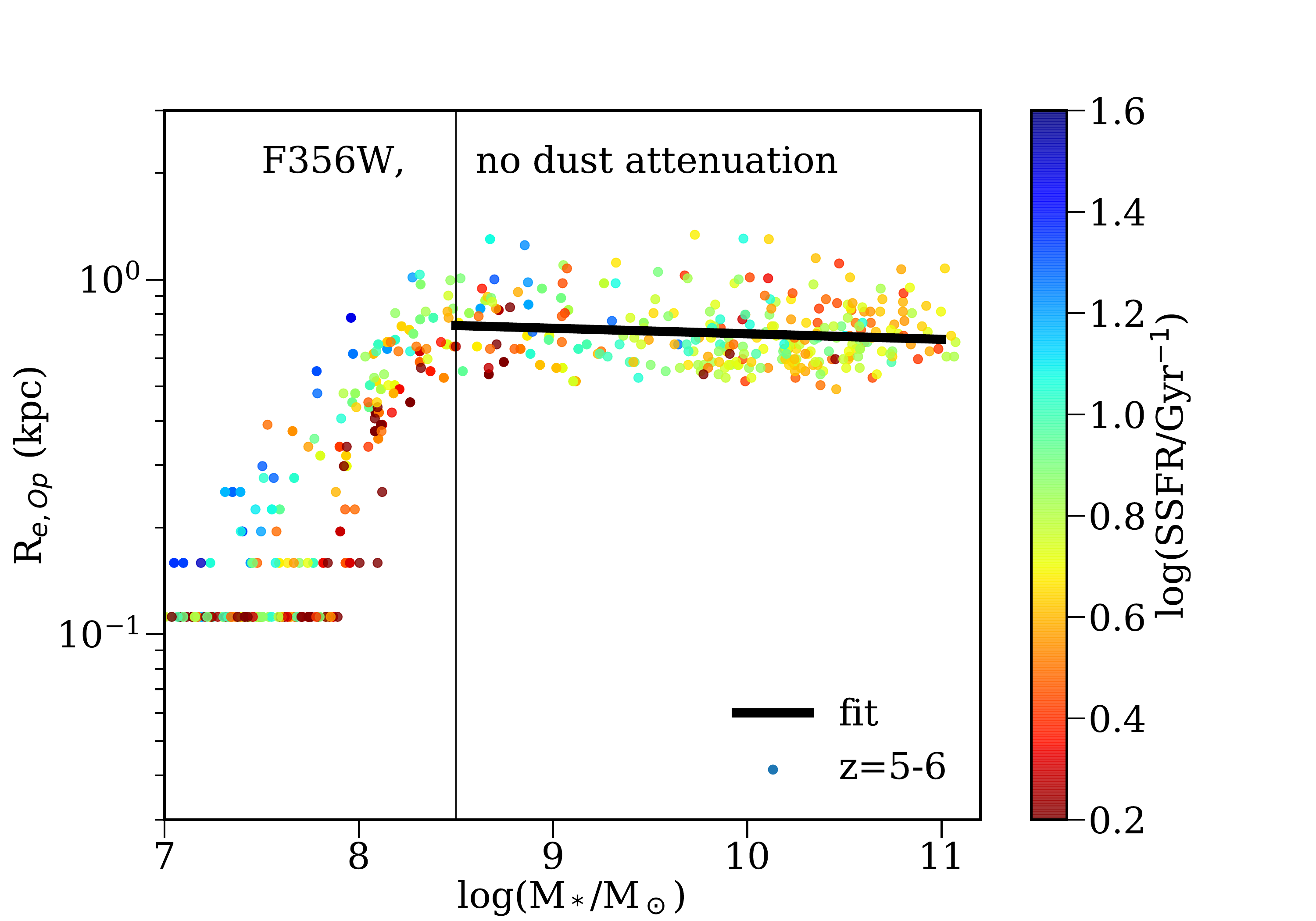}
   \includegraphics[width= \columnwidth]{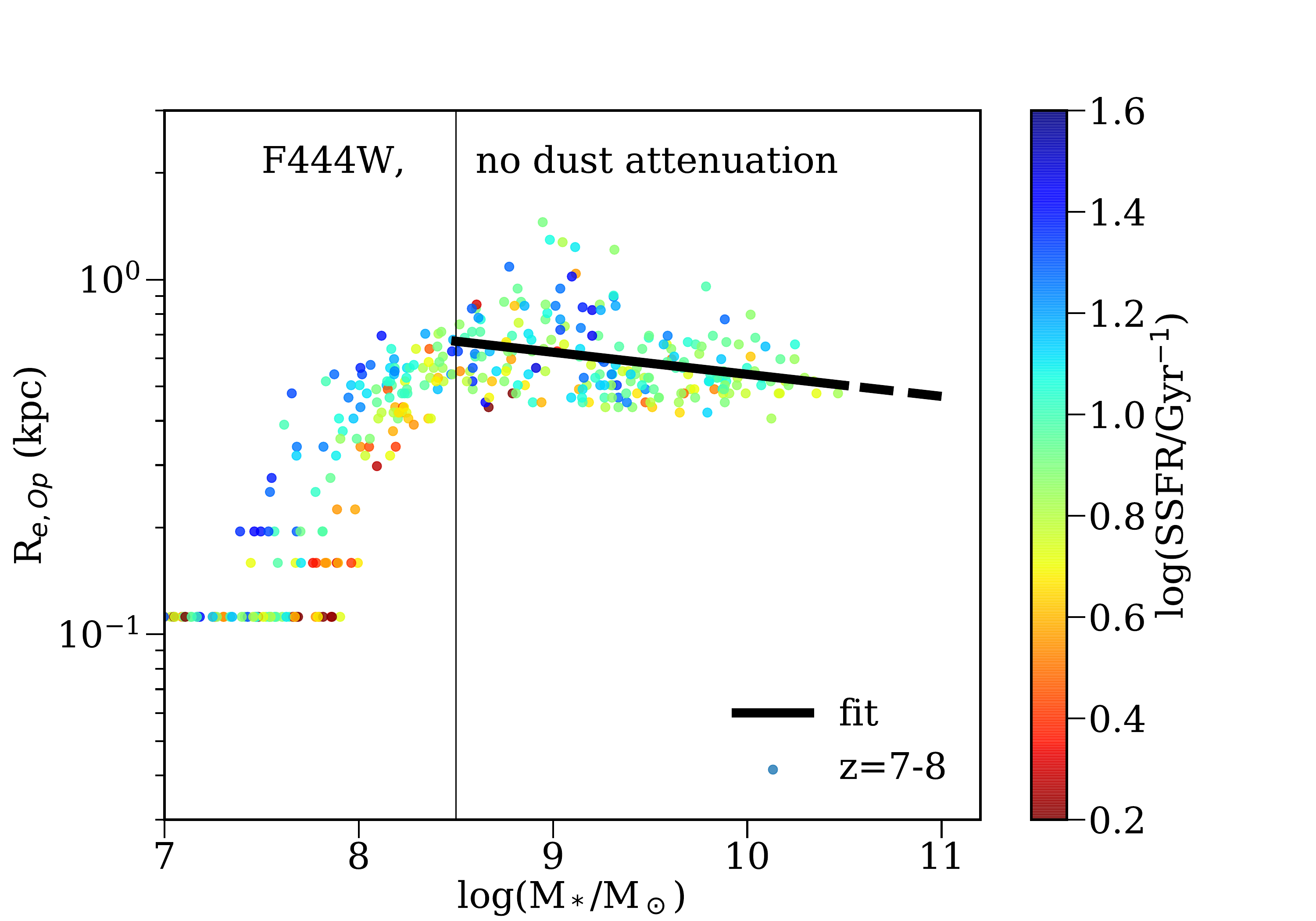}
   \caption{Size-Mass relation in the rest-frame optical without dust.  Points are coloured by the galaxy specific star-formation-rate (sSFR). Solid lines represent a fit to these points above $\Ms>5 \times 10^8 \ \msun$ (\protect\tab{fit}).  The solid vertical lines mark the turn-on mass, ${\rm M}_{*{\rm on}}$, for compaction at high-z \protect\citep{Cataldi25}. Due to this effect, the slope is negative above this mass. }
              \label{fig:sizeOp_Nodust}%
    \end{figure*}  

Galaxy sizes measured in the rest-frame optical also take into account stellar light coming from different stellar populations and they provide a more robust measurement of the actual size of galaxies.
\Fig{sizeOp} shows these relations for size measurements in F356W and F444W bands. 
Qualitatively, they are similar to the relations in the UV. 
Focusing on sizes in the F356W band ($z=5-6$), 
the steep slope at masses lower than $10^{8.5} \ \msun$ is also due to noise effects and incompleteness as in the UV case.
Above this stellar mass, the normalization, $\beta$, is higher than in the rest-frame UV case by a factor 1.4. This means that galaxies, in general, are more extended in the rest-frame optical compared to the UV at  $\Ms\simeq10^{9} \ \msun$.
At higher masses, the difference is much smaller due to the different slopes.
The optical slope is significantly shallower, $\alpha\simeq0.12$ (\tab{fit}) . This is consistent with the trend found by \cite{Allen25}. At longer wavelengths, size–mass relations become flatter. However, this value is smaller than in observations:  $\alpha=0.23 \pm 0.03$ \citep{Allen25} and $\alpha=0.19 \pm 0.07$ \citep{Danhaive25}. 
These differences are minor if we consider the high intrinsic scatter (0.2 dex) in the size measurements. Moreover, the FirstLight sample covers a range of stellar masses larger than  these observations.  The sizes of massive galaxies  ($\Ms>10^{10.5} \ \msun$) and  low-mass galaxies  ($\Ms<10^{9} \ \msun$)  favour this shallow slope.

The size-mass relation  in the rest-frame optical also evolves with redshift. The sizes measured in the F444W band ($z=7-8$) have a slightly lower normalization and shallower slopes. These sizes are consistent with observations at the same band and similar redshifts \citep{Ormerod24}. 
These shallow slopes suggest that more massive galaxies are also denser and this trend increases at higher redshifts. 
This is because
the slope is significantly shallower than $\alpha \simeq 0.5$, the slope required for a mass-independent stellar surface density, $\Sigma_* = 0.5 \Ms / ( \pi \Re^2)$. This value is not observed.
In fact, denser galaxies at higher masses are expected if compaction processes  are relevant at these scales.

\begin{table} 
\caption{Fit to the Size-mass relation  for $\Ms>5 \times 10^8 \ \msun$, \protect\equ{Refit}, at different redshifts, different bands and also without dust}
 \begin{center} 
 \begin{tabular}{cccc} \hline 
 redshift	&  Size in band     & $\alpha$	&  $\beta$   \\
 \hline
 5-6 		&		F115W		      &		$0.214 \pm 0.013$		&	$-0.22 \pm 0.13$	  \\
 7-9 		&		F150W		      &		$0.21	\pm 0.03$	&	$-0.3 \pm 0.3$  \\
 10-12	&		F200W		      &		$0.3	\pm 0.2$	&	$-0.6 \pm 2$   \\
 13-15	&		F277W		      &		0.21 (fixed)	&	$-0.76 \pm 0.09$  \\
 \hline 
 5-6		&		F356W			&	$0.122 \pm 0.008$	&	$-0.06 \pm 0.08$ \\
 7-8		&		F444W			&	$0.105 \pm 0.015$	&	$-0.11 \pm 0.14$ \\
  \hline 
 5-6		&		F356W w/o dust      &	$-0.016 \pm 0.003$	&	$-0.14 \pm 0.08$ \\
 7-8		&		F444W w/o dust	&	$-0.062 \pm 0.014$	&	$-0.20 \pm 0.13$ \\
  \hline 
 \end{tabular} 
 \end{center} 
  \label{tab:fit} 
 \end{table}     

   \begin{figure*}
   \centering
   \includegraphics[width= \columnwidth]{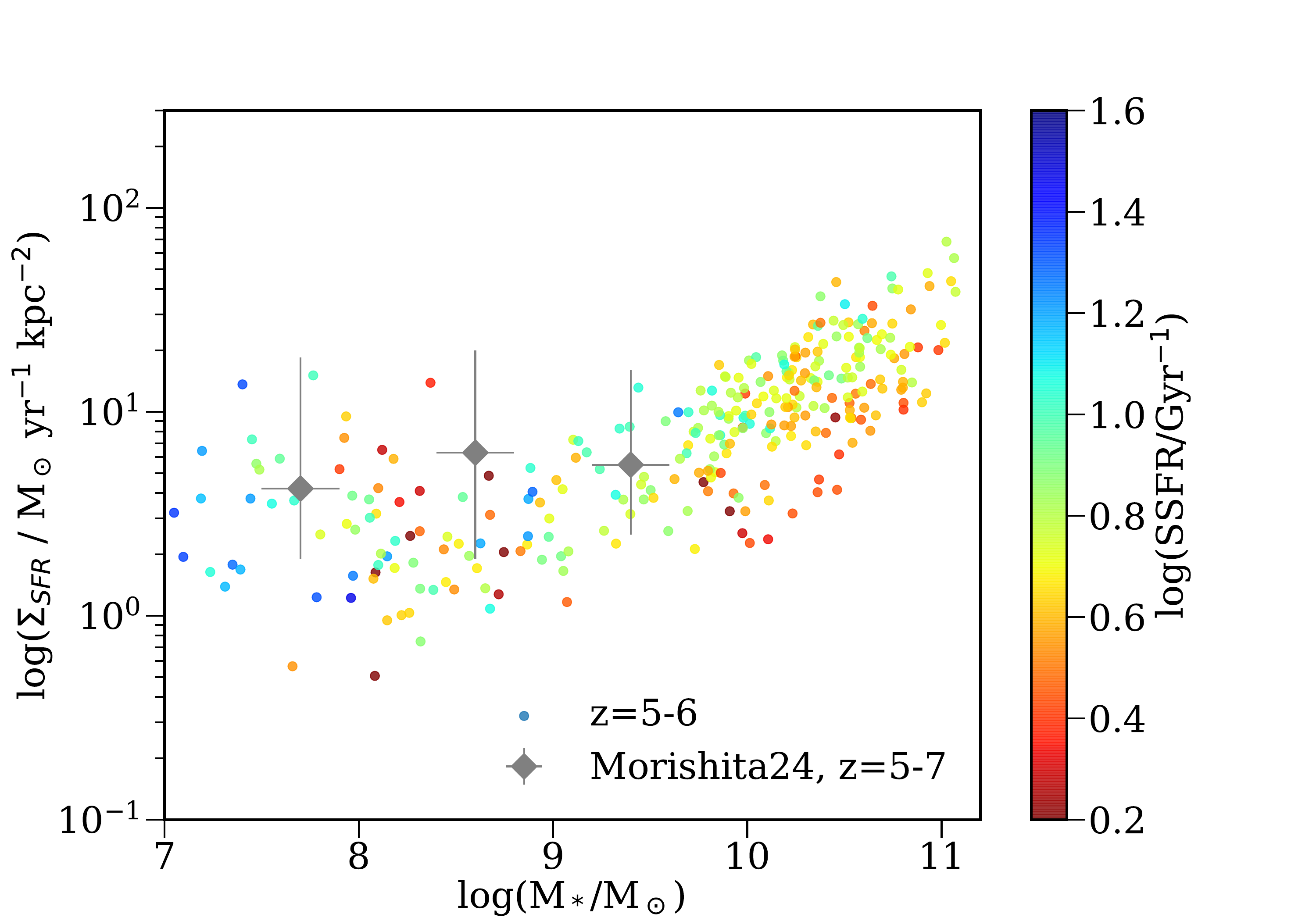}
   \includegraphics[width= \columnwidth]{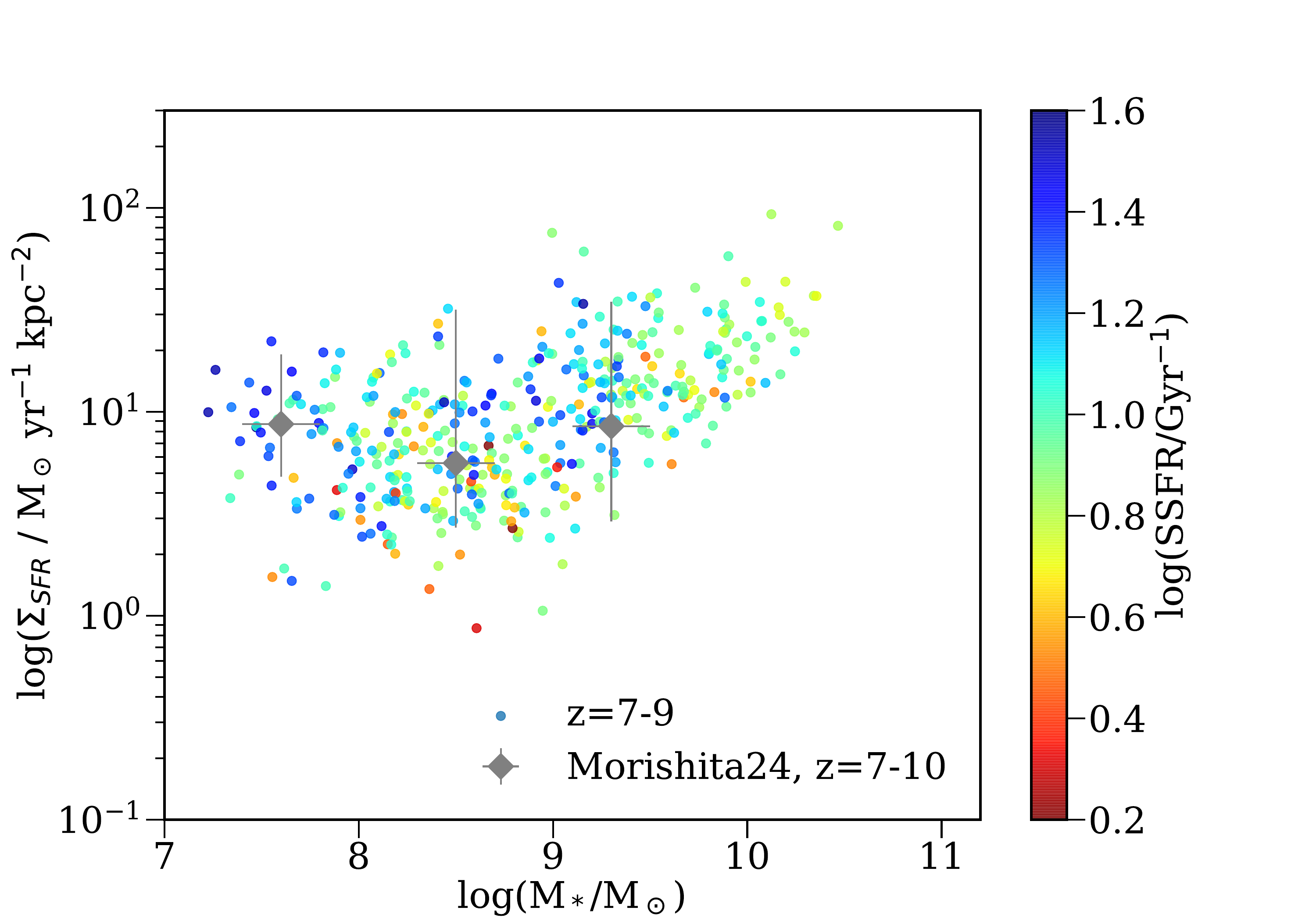}
    \includegraphics[width= \columnwidth]{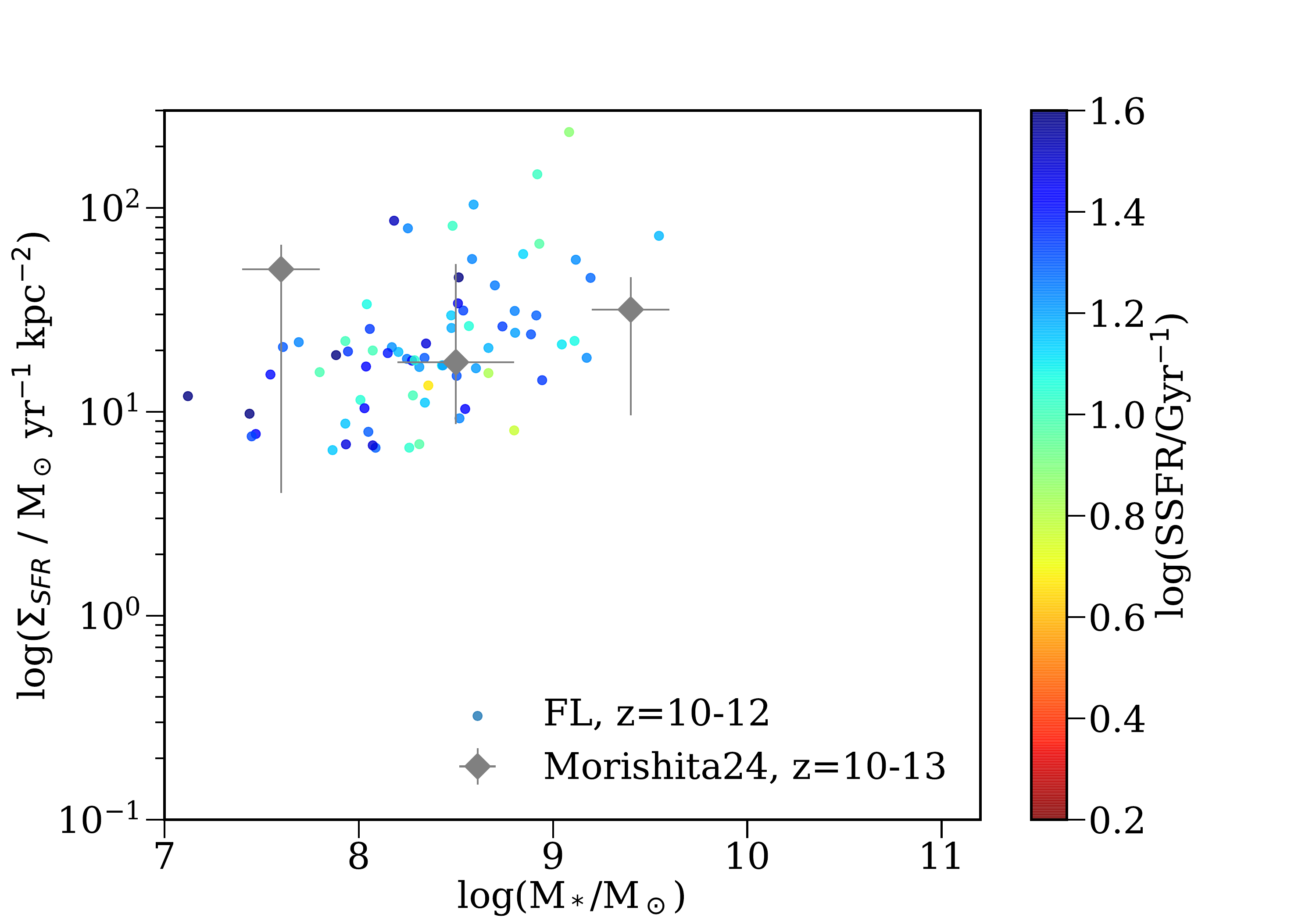}
   \includegraphics[width= \columnwidth]{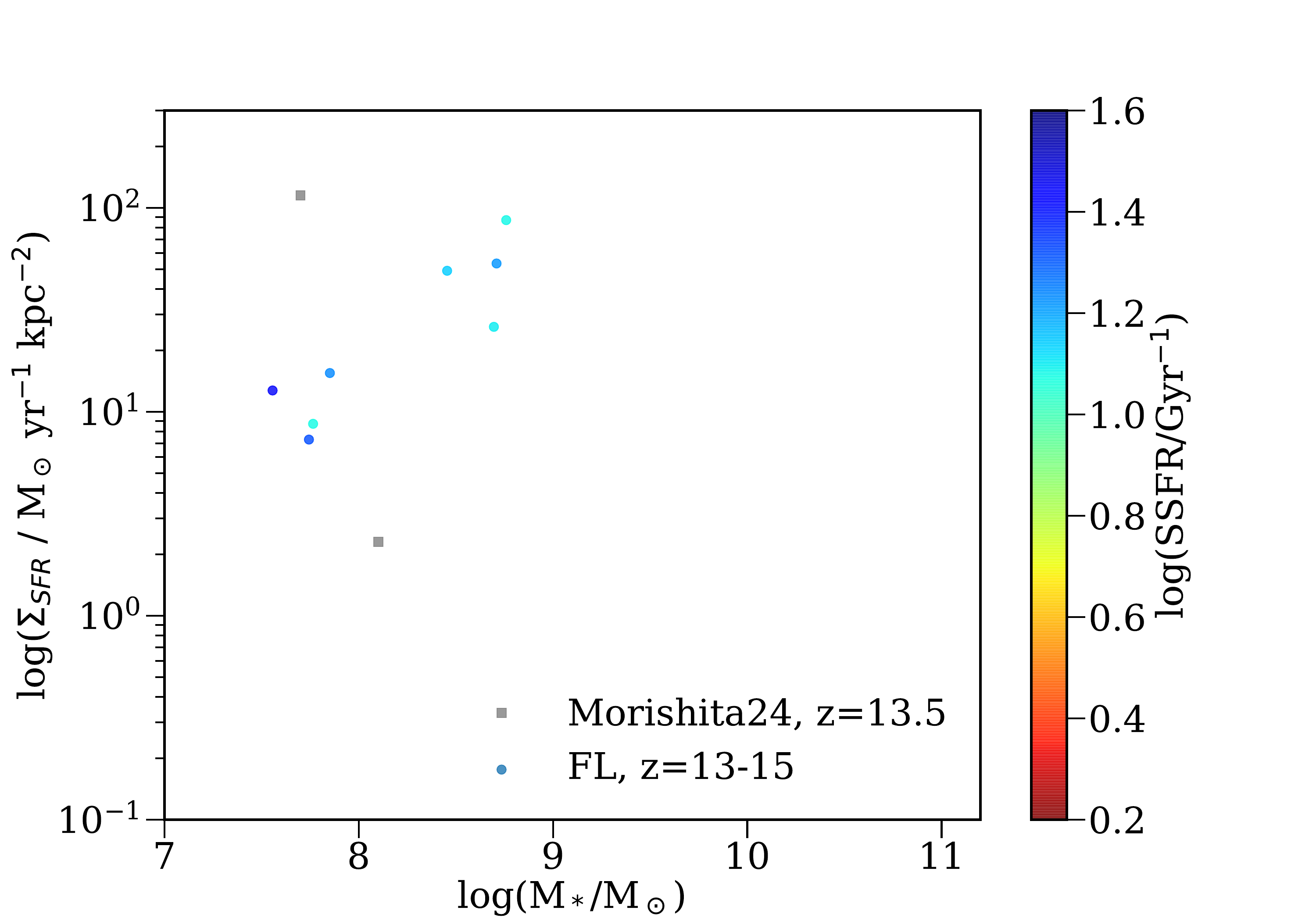}
   \caption{Star formation rate surface density versus stellar mass at different redshifts as in \protect\Fig{sizeUV}. Gray Points  describe median values from observations \protect\citep{Morishita24} at similar redshifts. Error bars in the y-axis mark the 25 and 75 percentiles. High densities  at high redshifts are expected due to the high galaxy efficiency at these times.}
              \label{fig:sigmaSFR}%
    \end{figure*}  
    
\subsection{Effect of dust attenuation}
\label{sec:dust}

In the previous section we found a shallow but positive slope for the size-mass relation in the rest-frame optical.     
However, dust attenuation may affect or change the  intrinsic relation of the stellar distribution within galaxies. 
\Fig{sizeOp_Nodust} shows the same relations based on images without dust.
Dust affects the size measurements in the rest-frame optical, especially above a mass of about $10^9 \ \msun$.

The most striking difference is the change of slope. The relations without dust show a slightly negative slope (\tab{fit}).
Massive galaxies are significantly smaller in images without dust attenuation due to  differential attenuation (\Fig{example1}). 
This makes galaxies more centrally concentrated in the images without dust especially in the rest-frame UV \citep{Wu20, Marshall22} but also in the rest-frame optical \citep{Roper22, Costantin23}.
An exception happens in the images generated using the Astrid simulation \citep{LaChance25}.  Their dust occlusion model does not significanly change the effective radius of their sample.
It seems that radiative transfer calculations are needed to properly recover the complex interactions between  light and dust at high redshifts.

The mass above which the slope becomes negative coincides with the turn-on mass, ${\rm M}_{*{\rm on}}$, 
above which compaction processes are efficient at high-z \citep{Cataldi25}. 
In this work, the half-mass radius, ${\rm r}^*_{\rm hm}$, shows a pronounced decrease with increasing mass, reaching ${\rm r}^*_{\rm hm} \simeq 100 \pc $ for galaxies above this threshold. In most cases, this is due to a strong compaction event. The origin of these events will be discussed further in a future paper (Cataldi et al in prep).
Regardless of its origin, fast inflows of gas into the galaxy center shrink galaxy sizes, even in the rest-frame optical.

\subsection{Evolution of the star-formation-rate surface density}
\label{sec:surfaceD}
 
 Wet compaction processes  and high galaxy efficiencies 
 lead to high values of star-formation-rate densities, defined as
 \begin{equation} 
 \Sigma_{\rm SFR} = 0.5 \  {\rm SFR} / ( \pi  \rm {R}_{\rm e, UV}^2) \ .
 \label{eq:SFRDen} 
 \end{equation}
 \Fig{sigmaSFR} shows these values as a function of mass at different redshifts. 
 The majority of points are relatively high, $\Sigma_{\rm SFR} > 1 \ \msun {\rm yr}^{-1} \kpc^{-2}$, in comparison with observations at lower redshifts, $z\leq3$ \citep{Skelton14, vanderWel14}. They are however comparable to  mean values at cosmic dawn \citep{Morishita24}. 
 These high values are due to a combination of high SFR \citep{PaperII} and smaller sizes in the rest-frame UV at cosmic dawn (\se{sizemassUV}).
 
 The surface densities remain independent of mass with a large scatter until $\sim10^9 \ \msun$ at $z=5-6$. Above this mass, the density increases with mass to extreme values, $\Sigma_{\rm SFR} \simeq 100 \ \msun {\rm yr}^{-1} \kpc^{-2}$ for very massive galaxies, $\Ms \sim10^{11} \ \msun$.
These massive starbursts harbour very high densities and therefore, stellar feedback is very inefficient, leading to a high, galaxy-averaged SF efficiency \citep{Dekel23, LiDekel24, PaperV}.   
They are observed in some examples at cosmic dawn \citep{Morishita24}.
AGN feedback, not included here, may limit these extreme values, if these massive galaxies also host active and massive black holes. 
 
   \begin{table} 
\caption{Fit to the redshift evolution of the effective radius in the rest-frame UV at  $\Ms= 10^9 \ \msun$, \protect\equ{Revo}, and comparison with previous simulations \protect\citep{Ceverino15} with a similar physical model as well as observations \protect\citep{Shibuya15, Morishita24}.}
 \begin{center} 
 \begin{tabular}{ccc} \hline 
  Fitting from	     & $\alpha_z$	&  $\beta_{z5}$   \\
 \hline
 This work	&	$-1.80 \pm 0.17$	&	$-0.12 \pm 0.17$ \\
 Ceverino et al. (2015)&  -1.31    &   -0.17 \\
  \hline
 Shibuya et al. (2015)  &  $-1.10 \pm0.06$		&	$-0.28 \pm 0.20$	\\
 Morishita et al. (2024)  & $-0.24 \pm 0.20$	&	$-0.33 \pm 0.18$      \\
   \hline 
 \end{tabular} 
 \end{center} 
  \label{tab:Revo} 
 \end{table}  
 
 The population of FirstLight galaxies with ${\rm log} (\Sigma_{\rm SFR} /  \msun {\rm yr}^{-1} \kpc^{-2})  \geq 1.5$ increases with redshift, with values comparable to observations at similar redshifts \citep{Morishita24, Zhang26}, as well as local starbursts \citep{Dopita02, KennicuttEvans12} and submilimiter galaxies \citep{Daddi10}. 
 At $z\simeq14$, $\Sigma_{\rm SFR} \simeq 100 \ \msun {\rm yr}^{-1} \kpc^{-2}$ are typical values even at relatively low masses, $\Ms\simeq10^9 \ \msun$.
 These extreme densities are similar to the values reported for JADESGDS-30934 at $z=13.8$ \citep{Morishita24} or beacon-1420+5253 4770 at $z=13.5$  \citep{Zhang26}. 
 The increase in $\Sigma_{\rm SFR}$ seen in FirstLight is due to an increase in the density of the SF gas and a higher galaxy efficiency \citep{PaperV}.  
 It is also possible that wet compaction processes are more efficient  at higher redshifts.  
 However, the connection between galaxy efficiency and compaction deserves a further study (Cataldi et al, in prep).   
    
\section{Discussion}
\label{sec:discussion}
    
\subsection{Accelerated evolution of the effective radius at high redshifts}
\label{sec:evo}

\begin{figure}
   \centering
   \includegraphics[width= \columnwidth]{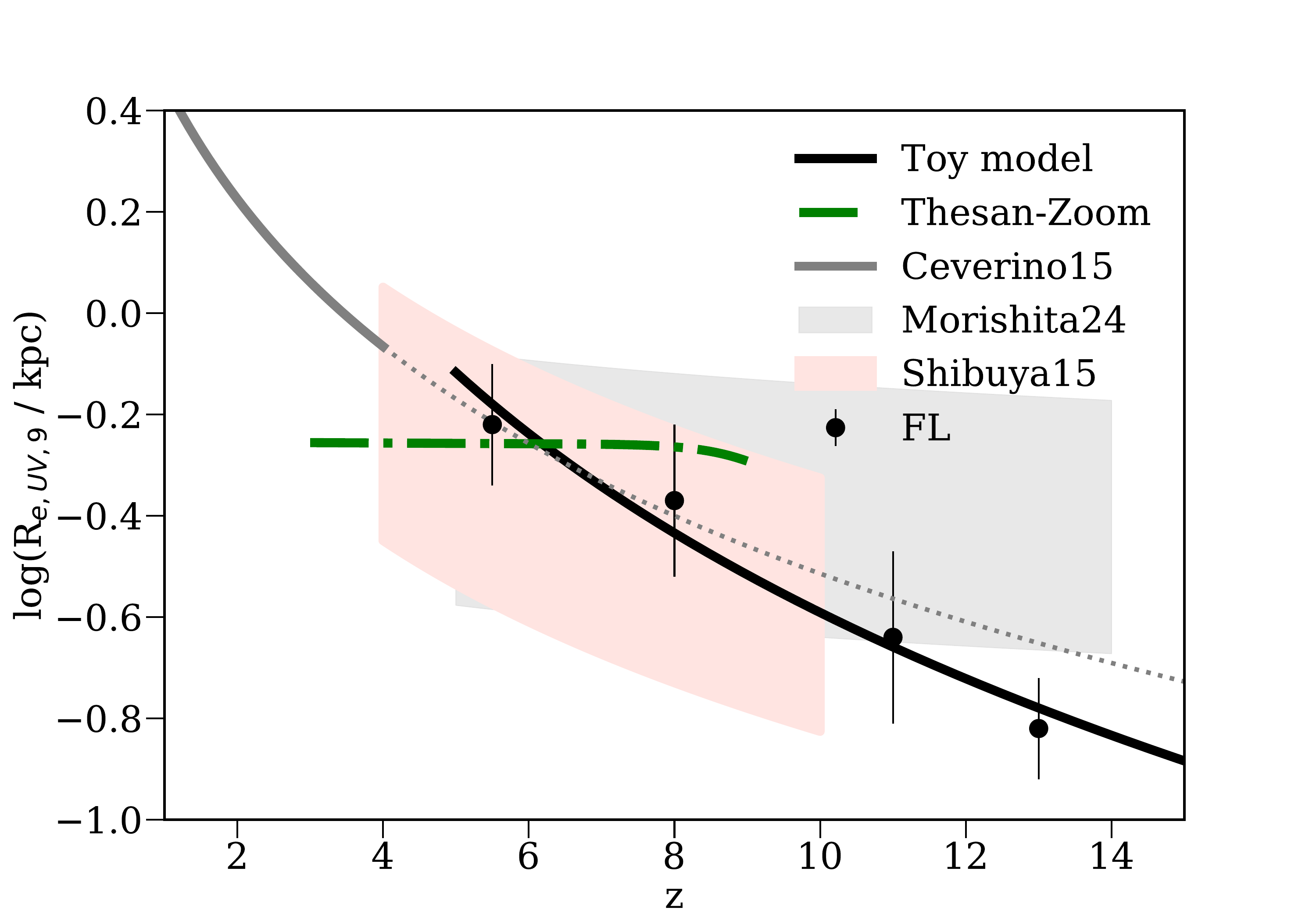}
   \caption{Evolution of the rest-frame UV effective radius with redshift at a fixed mass of $10^9 \ \msun$. Colour regions mark observations \protect\citep{Morishita24, Shibuya15}. The green line describes the double-power-law fit to the Thesan-Zoom simulations \protect\citep{McClymont25}. The grey line shows the fit of a previous simulation with a similar feedback model \protect\citep{Ceverino15}. FirstLight  (solid line) predicts a faster size evolution from $z\simeq5$ to  $z\simeq14$, driven by a high galaxy formation efficiency.}
              \label{fig:evoUV}%
    \end{figure}

There is a strong evolution of the size-mass relation at cosmic dawn. 
We can quantify this evolution by the redshift dependence of the effective radius in the rest-frame UV at a fixed mass of $10^9 \ \msun
$, \equ{Revo}.
This mass scale is not very sensitive to dust attenuation (\se{dust}) or to observational effects (Appendix).
\Fig{evoUV} compares this evolution with observations \citep{Shibuya15, Morishita24}. 
The normalization, $\beta_{z5}$, is consistent with observations, within the measured scatter.
However, the $z$-dependent slope is close to $\alpha_z\simeq -2$, much steeper than other simulations that show no evolution \citep{McClymont25} 
or JWST observations,  $\alpha_z\simeq -0.2$ (\tab{Revo}).
This is mostly driven by the fact that current observations lack statistics of objects at very high redshifts, $z>10$, 
and this epoch is key to observing this accelerated evolution.
It is possible that current observations miss small and faint galaxies at $z>10$.

Previous observations at lower redshifts ($z=1-4$) found slopes similar to $\alpha_z\simeq -1.2$ \citep{Mosleh12, Oesch10, Shibuya15}.  
This slope is close to the value of  $\alpha_z\simeq -1.3$ found in previous simulations that use a  feedback model similar to FirstLight \citep{Ceverino15}.
This value is higher than in \cite{Morishita24} but lower than in FirstLight. 
If the slopes at  high and low redshifts are consistent with each other, the galaxy size growth at a fixed mass accelerates during the first billion year in the history of the Universe with a z-dependence between $\alpha_z\simeq -1.6$ and $ \alpha_z\simeq  -2$. At lower redshifts, $z<5$, this dependence decreases to $\alpha_z \simeq -1$, only driven by the Universe expansion.

       \begin{figure}
   \centering
   \includegraphics[width= \columnwidth]{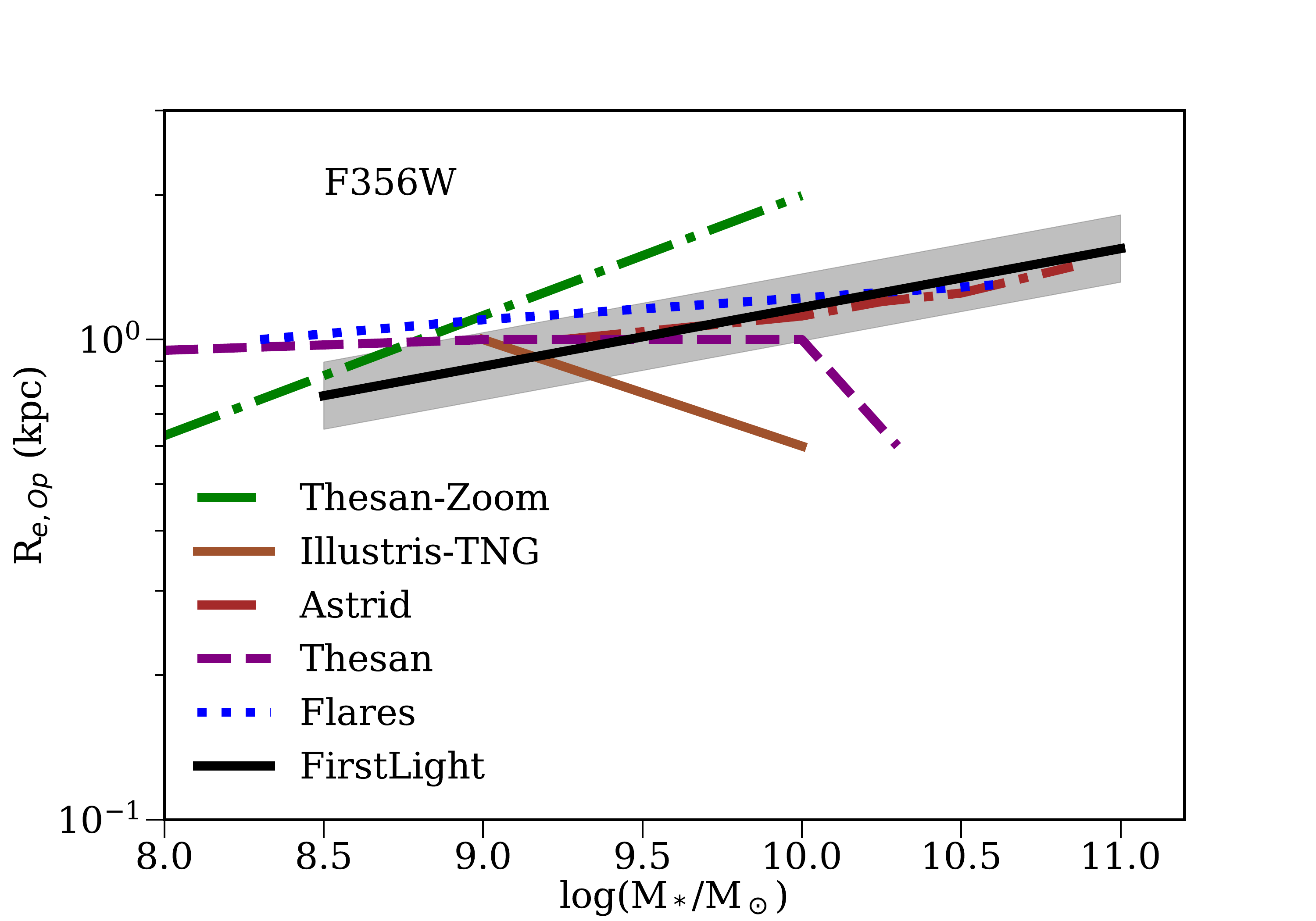}
     \caption{Size-Mass relation in the rest-frame optical at $z=5-6$ as in \Fig{sizeOp}.  The comparison with other simulations show flatter or negative slopes: Illustris-TNG \protect\citep{Costantin23}, Flares, \protect\citep{Roper22} Astrid \protect\citep{LaChance25}, Thesan \protect\citep{Shen24}. The Thesan-Zoom simulation shows a much steeper slope \protect\citep{McClymont25}. The grey region marks the 25\% and 75\% percentiles of the FirstLight scatter. }
              \label{fig:compSim}%
    \end{figure} 

The accelerated evolution of the galaxy size at a fixed mass can be understood by a simple analytical model, inspired by previous toy models \citep{Dekel13, Ceverino15}. 
We first relate the galaxy effective radius with the halo virial radius:
\begin{equation} 
\Re \propto \tilde{\lambda} \  {\rm R}_{\rm vir},
\label{eq:Rtoy}
  \end{equation}
where the factor $ \tilde{\lambda}$ encapsulates all processes related to the change of angular momentum during gas accretion into the halo center \citep{Danovich15}. A fit to previous simulations at high redshifts found a minor redshift dependence \citep{Ceverino15}. At a fixed halo mass, angular momentum is lost more efficiently at higher redshifts:    
  \begin{equation} 
  \tilde{\lambda} \propto (1+z)_{z5}^{-0.3}.
  \label{eq:lambdatoy}
    \end{equation}
The virial radius scales with redshift and halo mass as     
 \begin{equation} 
  {\rm R}_{\rm vir} \propto \ {\rm M}_{11}^{1/3} (1+z)_{z5}^{-1},
\label{eq:Rhalotoy}  
   \end{equation}
where  ${\rm M}_{11}={\rm M}_{\rm h} / 10^{11} \ \msun$ is the rescaled halo mass and it is proportional to:
   \begin{equation} 
   {\rm M}_{11} \propto     {\rm M}_{*,9} \  \epsilon_{*}^{-1},
 \label{eq:Mhalotoy}    
     \end{equation}  
where  ${\rm M}_{*,9}=  {\rm M}_* / 10^9 \ \msun$ is the rescaled stellar mass and $ \epsilon_{*}$ is the integrated galaxy efficiency. At a fixed stellar mass of $10^9 \ \msun$, it scales as 
    \begin{equation} 
         \epsilon_{*,9} \propto (1+z)_{z5}^{1.4}  ,
         \label{eq:epsilontoy}
      \end{equation}          
according to the evolution of the integrated galaxy efficiency of galaxies at this mass scale at $z\geq5$ \citep{PaperV}. 
At lower redshifts, $z=1-4$, simulations with the same feedback model show little evolution with redshift \citep{Ceverino23}.       
We  use \equ{Rtoy}, \equ{Rhalotoy}, and \equ{Mhalotoy}  to write the explicit dependence of the effective radius as
    \begin{equation} 
     \Re \propto    \tilde{\lambda}  \ \epsilon_{*}^{-1/3} \ (1+z)_{z5}^{-1} \  {\rm M}_{*,9}^{1/3}.
     \label{eq:Reff}
      \end{equation}        
Finally, using the redshift dependences in \equ{lambdatoy} and \equ{epsilontoy}, we  write the redshift evolution of the effective radius of galaxies at this fixed mass of $10^9 \ \msun$ as      
   \begin{equation} 
     R_{\rm e,9} \propto   (1+z)_{z5}^{-1.8}
     \end{equation} 
This model yields $\alpha_z < -1$ at $z\ge5$ due to the strong evolution of the galaxy efficiency and a significant loss of angular momentum at these early times.   
This simple toy model is able to reproduce the accelerated evolution seen in FirstLight $z\geq5$. At lower redshifts,  the size evolution gets closer to   $\alpha_z \simeq -1$ because the evolution of the galaxy efficiency slows down.

\subsection{Comparison with other simulations}
    
    Previous simulations tend to produce size-mass relations with little mass dependence. \Fig{compSim} compares the size-mass relations at $z=5-6$ from different models.
        At low masses, $\Ms \simeq10^9 \ \msun$, other simulations \citep{Roper22, Shen24} show relatively flatter relations compared to  FirstLight. 
        This could be related to resolution effects.
    Thesan-Zoom \citep{McClymont25} is the only simulation with a relatively similar mass resolution. Its dark-matter particle mass is 5 times higher than FirstLight for this mass-scale. It shows a significantly positive trend, steeper than FirstLight or JWST observations.
    This could be due to their strong feedback model that prevents a high galaxy efficiency at high-$z$. Most probably, the density of SF gas is lower than in FirstLight and therefore their massive galaxies are less concentrated. 
    
    At high masses, $\Ms \geq 10^{10} \ \msun$, dust attenuation shapes the relation. FirstLight is consistent with the results from Astrid \citep{LaChance25} and Flares \citep{Roper22}, although they have much lower resolution and different dust implementation than FirstLight.
    Thesan \citep{Shen24} and TNG \citep{Costantin23} show negative slopes. 
    Both simulations employ the same Illustris-TNG model  \citep{Pillepich18, Nelson19}. \cite{Costantin23} report a transition from negative to positive slopes between $z = 5$ and $z = 4$, which is not observed.
This negative slope could be due to little dust content in these simulations at high-z due to very strong galaxy outflows from stellar and AGN feedback. 
    
%
%

\section{Summary and conclusions}
\label{sec:conclusions}
    
    We use the FirstLight database of 430 zoom-in cosmological simulations of a mass-complete sample of galaxies with a spatial resolution of about $10-40$ parsecs in a $(120 \Mpc)^3$ comoving volume. We generate synthetic images in seven JWST bands to understand size evolution at cosmic dawn. 
 The main highlights of this paper are the following:
   \begin{enumerate}
      \item The size-mass relation is already in place at $z\simeq14$ and its mass-dependent slope does not evolve. 
   \item {There is a large diversity of galaxy sizes at a fixed mass. 
    Extended (compact) galaxies tend to have higher (lower)  sSFR.} 
      \item At a fixed mass, galaxy size evolves very fast, as the normalization increases by 0.5 dex between $z\simeq14$ and $z\simeq6$, in about 600 Myr.
      \item Differential dust attenuation makes larger sizes and modifies the size-mass slope even in the rest-frame optical.
      \item The SFR surface density increases with redshift.
    \end{enumerate}     
    
FirstLight reveals an accelerated size growth at $z\geq5$, mostly driven by the evolution of the galaxy efficiency, described as the stellar-to-halo mass ratio.
At lower redshifts, this evolution slows down.
However, FirstLight does not reach these redshifts.
We can rely on the VELA-6 simulations \citep{Ceverino23} that use the same feedback model.
In these simulations, the galaxy efficiency evolves faster at earlier times, $z=4-6$, than at later times, $z=1-4$.
Then, we expect that size evolution slows down at later times, according to \equ{Reff}.    
Future analysis at lower redshifts can confirm this trend.
    

The predictions of galaxy sizes require a multi-layer, multi-scale approach.
The next generation of cosmological simulations probably requires parsec-scale resolution 
in order to properly resolve the distribution of dense gas in galaxies, star formation and
self-regulation at stellar birth, including key processes such as turbulence, feedback from black holes, as well as magnetic fields
and cosmic rays. 
In addition, more accurate models of dust production, growth and destruction at  these early times are needed to recover the complex mix of dust and stars in these first galaxies \citep{Nakazato26}. This is crucial for a more accurate radiative transfer calculations and the generation of synthetic images that can be compared with future observations at different wavelengths, using JWST but also ALMA and the next-generation of ground-based telescopes.


\begin{acknowledgements}
We acknowledge stimulating discussions with Claudio dalla Vecchia,  Arianna di Cintio, Enrico Garaldi, Javier Alvarez-Marquez, Pablo Perez-Gonzalez, Leonardo Pellizza, Susana Pedrosa and Pedro Cataldi. 
     The toy model in section \se{evo} is a tribute to Avishai Dekel.
     The authors gratefully acknowledge the Gauss Center for Supercomputing for funding this project by providing computing time on the GCS Supercomputer SuperMUC at Leibniz Supercomputing Centre (Project ID: pr92za). 
     The authors thankfully acknowledge the computer resources at MareNostrum and the technical support provided by the Barcelona Supercomputing Center (RES-AECT-2020-3-0019).
     This work used the v2.1 of the Binary Population and Spectral Synthesis (BPASS) models as last described in Eldridge et al. (2017).
     DC is supported by research grant PID2021-122603NB-C21 funded by the Ministerio de Ciencia, Innovaci\'{o}n y Universidades (MI-CIU/FEDER), project  PID2024-156100NB-C21  financed by MICIU/AEI /10.13039/501100011033 / FEDER, EU., and the research grant CNS2024-154550 funded by MI-CIU/AEI/10.13039/501100011033.
   YN acknowledges support from the Flatiron Research Fellowship. The Flatiron Institute is a division of the Simons Foundation.
     RSK and SCOG acknowledge funding from the ERC via Synergy Grant "ECOGAL" (project ID 855130), from the German Excellence Strategy via the Heidelberg Cluster of Excellence (EXC 2181 - 390900948) "STRUCTURES", and from the German Ministry for Economic Affairs and Climate Action in project ``MAINN'' (funding ID 50OO2206). RSK and SCOG also thank for computing resources provided by the Ministry of Science, Research and the Arts (MWK) of the State of Baden-W\"{u}rttemberg through bwHPC and DFG through grant INST 35/1134-1 FUGG and for data storage at SDS@hd through grant INST 35/1314-1 FUGG.
     The project that gave rise to these results received the support of a fellowship from the “la Caixa” Foundation (ID 100010434). The fellowship code is LCF/BQ/PR24/12050015. LC acknowledges support from grants PID2022-139567NB-I00 and PIB2021-127718NB-I00 funded by the Spanish Ministry of Science and Innovation/State Agency of Research  MCIN/AEI/10.13039/501100011033 and by “ERDF A way of making Europe”.
\end{acknowledgements}

%
 \bibliography{Size_v4} 
%

\appendix

\section{Effect of the background noise}    

\begin{figure}
   \centering
   \includegraphics[width= \columnwidth]{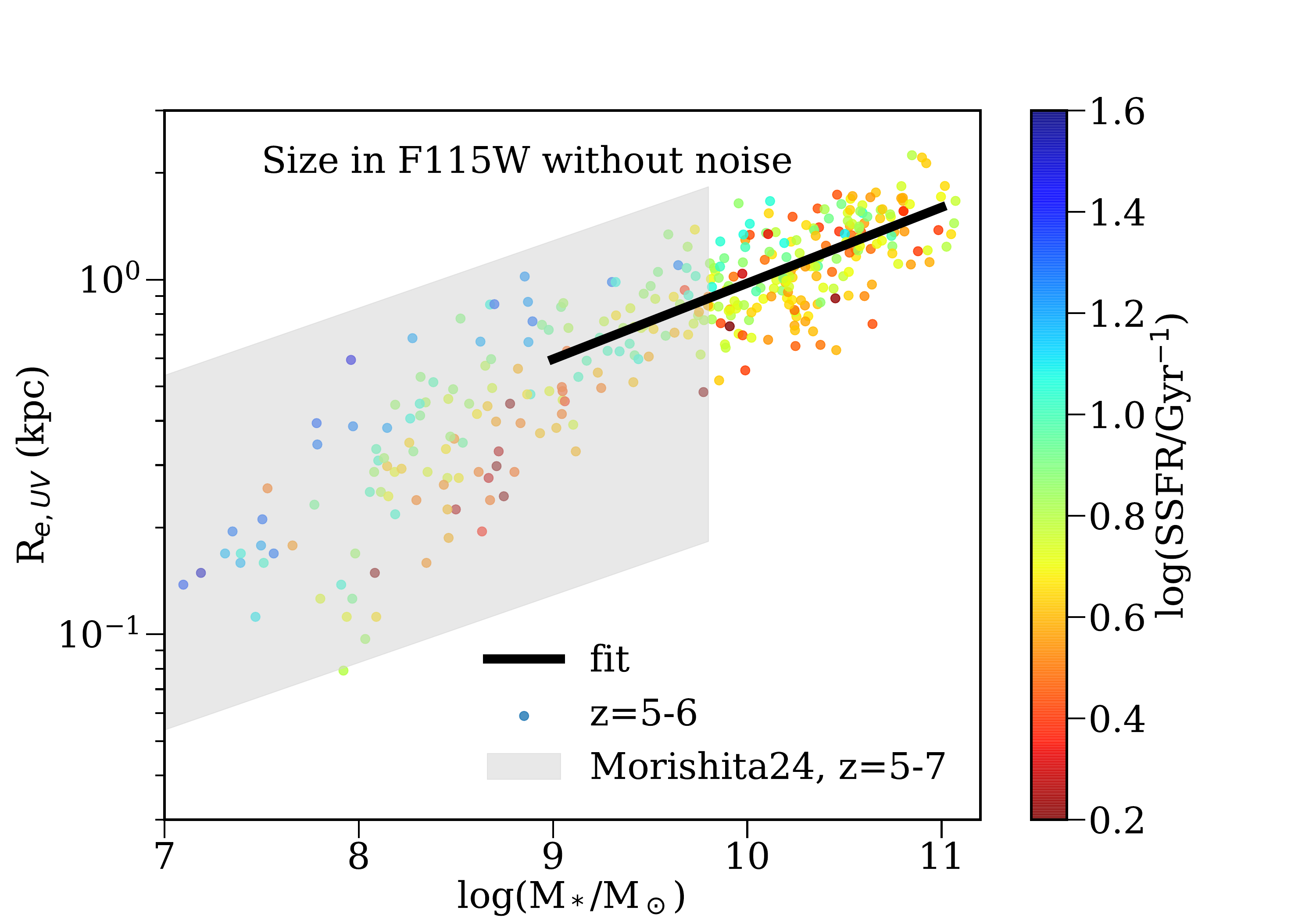}
   \caption{The effect of the background noise in the size-mass relation in the rest-frame UV at z=5-6, as in \protect\Fig{sizeUV}.}
              \label{fig:noise}%
    \end{figure}

The background noise can affect the estimation of the galaxy size in multiple ways. 
For example, regions of galaxies with very low surface brightness can be missed if  they fall below the background noise.
\Fig{noise} shows the size-mass relation in the rest-frame UV at $z=5-6$, as in \Fig{sizeUV}, without this noise. 
We still measure the half-light radius using all pixels above 0.008 ${\rm MJy \ sr}^{-1}$, the same threshold used in previous figures. 
This allows an easy comparison, especially at low masses, $\Ms \leq 10^{8.5} \ \msun$, where differences are more noticeable.

The reported break in the relation is less evident. 
Instead, there is an increase in the scatter, driven by the bursty SF history of these low-mass galaxies, which also broadens the star formation main sequence \citep{PaperII}.
Interestingly, the addition of noise includes galaxies with low surface brightness into our sample.
These galaxies have a very low effective radius, $\Re \leq 300 \pc$, and they are just below our detection threshold in the images without noise.

\end{document}